\documentclass{emulateapj}
\usepackage{graphicx}
\usepackage[utf8]{inputenc}
\usepackage[flushleft,online,para]{threeparttable}

\newcommand{\multifast}{multi-EXOFAST }
\newcommand{\Multifast}{Multi-EXOFAST }
\newcommand{\degrees}{\ensuremath{^{\circ}}}

\newcommand{\arcs}{\mbox{\ensuremath{^{\!\prime\prime}}}}
\newcommand{\be}{\begin{equation}}
\newcommand{\ee}{\end{equation}}
\newcommand{\bea}{\begin{eqnarray}}
\newcommand{\eea}{\end{eqnarray}}
\newcommand{\mjup}{{\rm M}_{\rm J}}
\newcommand{\rjup}{{\rm R}_{\rm J}}
\newcommand{\teff}{\ensuremath{T_{\rm eff}}}

\newcommand{\feh}{{\rm [Fe/H]}}

\newcommand{\loggp}{\ensuremath{\log{g_{\rm P}}}}
\newcommand{\loggstar}{\ensuremath{\log{g_\star}}}

\newcommand{\chisq}{\ensuremath{\chi^{\,2}}}

\newcommand{\chisqred}{\ensuremath{\chi^{\,2}_{red}}}
\newcommand{\vsini}{\ensuremath{\,{v\sin{i_\star}}}}
\newcommand{\bjdtdb}{\ensuremath{\rm {BJD_{TDB}}}}

\newcommand{\msun}{\ensuremath{\,{\rm M}_\Sun}}
\newcommand{\rsun}{\ensuremath{\,{\rm R}_\Sun}}
\newcommand{\lsun}{\ensuremath{\,{\rm L}_\Sun}}
\newcommand{\mj}{\ensuremath{\,{\rm M}_{\rm J}}}
\newcommand{\rj}{\ensuremath{\,{\rm R}_{\rm J}}}
\newcommand{\fave}{\langle F \rangle}
\newcommand{\fluxcgs}{10$^9$ erg s$^{-1}$ cm$^{-2}$}

\newcommand{\actaa}{Acta Astronomica}

\usepackage{color}
\usepackage[dvipsnames]{xcolor}
\usepackage[normalem]{ulem}

\begin{document}

\title{Transit Timing Variation Measurements of WASP-12\MakeLowercase{b} and Qatar-1\MakeLowercase{b}: \\ No Evidence for Additional Planets}

\author{
Karen A.\ Collins\altaffilmark{1,2,3}, 
John F.\ Kielkopf\altaffilmark{3}, and
Keivan G.\ Stassun\altaffilmark{1,2}
}
\altaffiltext{1}{Department of Physics \& Astronomy, Vanderbilt University, Nashville, TN 37235, USA}
\altaffiltext{2}{Department of Physics, Fisk University, Nashville, TN 37208, USA}
\altaffiltext{3}{Department of Physics \& Astronomy, University of Louisville, Louisville, KY 40292, USA}

\shorttitle{WASP-12b and Qatar-1b TTVs}

\begin{abstract}

WASP-12b and Qatar-1b are transiting Hot Jupiters for which previous works have suggested the presence of transit timing variations (TTVs) indicative of additional bodies in these systems---an Earth-mass planet in WASP-12 and a brown-dwarf mass object in Qatar-1. Here,
we present 23 new WASP-12b and 18 new Qatar-1b complete (or nearly complete) transit observations. We perform global system fits to all of our lights curves for each system, plus RV and stellar spectroscopic parameters from the literature. The global fits provide refined system parameters and uncertainties for each system, including precise transit center times for each transit. The transit model residuals of the combined and five minute binned light curves have a RMS of 183 and 255 parts per million (ppm) for WASP-12b and Qatar-1b, respectively. Most WASP-12b system parameter values from this work are consistent with values from previous studies, but have $\sim40-50$\% smaller uncertainties. Most of the Qatar-1b system parameter values and uncertainties from this work are consistent with values recently reported in the literature. We find no convincing evidence for sinusoidal TTVs with a semi-amplitude of more than $\sim35$~s and $\sim25$~s in the WASP-12b and Qatar-1b systems, respectively. 
\end{abstract}


\section{Introduction}
\label{sec:intro}

The search for Earth-like planets has been underway from the ground since well before the \textit{Kepler} mission \citep{boru10} discovered thousands of extra-solar planets, including several Earth-sized planets and smaller. One method used by ground-based observers enables the discovery of additional planets in systems with at least one transiting planet through high precision measurements of transit timing variations (TTVs; \citealt{agol2005,holman2005}). \citet{agol2005} show that an Earth-mass planet in 2:1 resonance with a 3-day period transiting hot Jupiter would cause timing variations of $\sim3$ minutes, which would be accumulated over a year. Ground-based half-meter class telescopes can be used to produce transit timing data with better than one minute precision for most planets with host stars brighter than $V\sim13$.

To properly characterize a TTV signal, good phase coverage and a long baseline of measurements are needed. Ideally, nearly all transits during the baseline period should be observed. This is of course nearly impossible from the ground due to the Earth's rotation and weather. \citet{veras2011} suggests that in most cases at least 50 consecutive transit observations are necessary to have a reasonable chance of characterizing the perturbing planet and its orbit.

Even if a TTV curve is well sampled, it is notoriously difficult to uniquely infer the perturbing planet's mass and orbital parameters strictly from the TTV data \citep{ford2007,nesvorny2008,payne2010,boue2012,lithwick2012}. Degeneracies exist in orbital period, eccentricity, and inclination. RV data, transit duration variations, or systems with multiple transiting planets can help to break the degeneracies. 

So far, no convincing TTVs have been reported from ground-based measurements. However, several authors have claimed evidence of low-amplitude periodic TTVs, followed by other authors with differing conclusions. For example, analysis of WASP-3b TTVs by multiple authors has produced differing conclusions. \citet{maciejewski2010} analyzed 13 WASP-3b transits and reported a sinusodal TTV with period $\sim127$ d and peak value $\sim2$  min. Then \citet{eibe2012} found a lower significance in the TTV detection, but a possible detection of transit duration variations. \citet{montalto2012} studied 38 WASP-3b light curves and found no evidence of periodicity in the TTVs. Small amplitude TTV claims are inconsistent across ground-based timing measurement analyses and, so far, no perturber planet has been confirmed from them.

On the other hand, the almost continuous Kepler light curves have demonstrated a number of clear TTV detections (e.g. \citealt{holman2010, lissauer2011, ballard2011, steffen2013, yang2013}). The Kepler TTV data are responsible for the discovery of non-transiting planets and provide confirmation of the planetary nature of the transiting planet(s) and estimates of their masses, without the necessity of RV data.

\citet{steffen2012} searched six quarters of Kepler data for planetary companions orbiting near hot Jupiter planet candidates ($1\le P \le5$ d) by looking for multiple transiting planets and evidence of TTVs. Special emphasis was given to companions between the 2:1 interior and exterior mean-motion resonances. They found no evidence for nearby companion planets to 63 hot Jupiter candidates. However, five out of 31 warm Jupiter systems ($6.3\le P \le15.8$ d) do show multiple transiting planets and/or evidence of TTVs.

Nevertheless, using K2 data \citep{howell2014}, \citet{becker2015} recently discovered two additional planets (WASP-47c, WASP-47d) transiting the ground-based discovered system hosting the hot-Jupiter planet WASP-47b \citep{hellier2012}. The K2 data show  WASP-47b transit timing variations with a period of $\sim 52$ d and peak variation of $\sim 45$ s.

WASP-12b and Qatar-1b are transiting Hot Jupiters for which previous works have suggested the presence of transit timing variations (TTVs) indicative of additional bodies in these systems---an Earth-mass planet in WASP-12 (\citealt{maciejewski2013}; M13) and a brown-dwarf mass object in Qatar-1 (\citealt{essen2013}; E13).
In this work, we present high-precision photometric observations of 23 full WASP-12b and 18 full Qatar-1b transits. We combine all transits from each system with spectroscopic and RV data from the literature and perform a global system fit for each system. The results provide refined system parameters for WASP-12b and Qatar-1b and high-precision transit center times for each transit observation.   

We describe the WASP-12 and Qatar-1 systems and the previous reports of TTV measurements in \S \ref{sec:prev}.
We describe the data that we utilize, including data from the literature and our extensive new transit observations, in \S \ref{sec:data}. We discuss the methods used for careful time keeping in \S \ref{sec:timing}, data reduction in \S \ref{sec:wasp12dr}, and for global fitting in \S \ref{sec:globalfit}. We present the results of this study in \S \ref{sec:results}, including the global fits, improved system parameter results, and TTV results.  
We discuss our TTV results in the context of previous reports in \S \ref{sec:disc},
and summarize our conclusions in \S \ref{sec:wasp12conclusions}.

\section{WASP-12\MakeLowercase{b} and Qatar-1\MakeLowercase{b}: Previous Transit Timing Variation Reports and Limitations}\label{sec:prev}

WASP-12b is a hot Jupiter exoplanet discovered by the WASP survey \citep{pollacco2006} and announced by \citealt{hebb2009} (H09). It orbits a $V\sim11.7$ evolved late-F star and has an orbital period of $P=1.09$~d, a semi-major axis of only 3.1 stellar radii, and a highly inflated radius $R_P=1.8~\rjup$. \citet{bergfors2013} reported a faint ($\Delta i^{\prime}=4.03$) elongated object $\sim1\arcs$ from WASP-12. \citet{crossfield2012} confirmed the detection. Using the Keck telescope, \citet{bechter2014} resolved the neighboring object and confirmed it to be a binary composed of two M3V stars that are orbiting WASP-12 as part of a hierarchical triple star system. The projected separation of the binary from WASP-12 is $\sim300$~AU, corresponding to a period of several thousand years, so any influence on the orbital dynamics of WASP-12b should be negligible.  

Qatar-1b is a hot Jupiter exoplanet and is the first planet discovered by the Qatar Exoplanet Survey \citep{alsubai2013}. The planet was announced by \citealt{al11} (A11). The host star is a $V\sim12.8$ metal-rich K-dwarf star, and the planet has a circular orbital with period $P=1.42$~d, a semi-major axis of $\sim6$ stellar radii, a mass $M_P=1.09~\mjup$ and a radius $R_P=1.2~\rjup$. 

The short orbital periods of WASP-12b and Qatar-1b result in relatively frequent opportunities to observe complete transit events from the ground, which has prompted other groups to conduct detailed studies of the systems. M13 acquired 61 partial or complete transit light curves from 2009 to 2012 from 14 telescopes distributed around the world. They classified 19 of the transits as high quality based on the RMS of the model residuals and transit coverage of at least 75\%. Out of the 19 high quality transits, 11 have both ingress and egress coverage and baseline data before and after the transit. Data reduction and differential photometry were performed using differing methods, depending on the originating observatory. Detrending was implemented by fitting a second order polynomial and a fixed WASP-12b transit model from \citet{maciejewski2011}, which was based on one transit. M13 find tentative evidence for a transit timing variation (TTV) signal which has a period of $545\pm22$~d and a semi-amplitude of $59\pm11$~s, and they suggest that the possible perturbing body has a mass of $0.1~\mjup$ and a 3.6 day eccentric orbit.

\citealt{covino2013} (C13) obtained HARPS-N in-transit high-precision radial velocities to measure the Rossiter–McLaughlin (RM) effect in the Qatar-1b system, and out-of-transit measurements to redetermine the spectroscopic orbit. They found that the orbit is consistent with circular and has a well-aligned spin-orbit angle of $\lambda=-8.4\pm7.1\degrees$. E13 presented a detailed TTV analysis of 26 Qatar-1b transits covering a baseline of 18 months and found evidence for a 190 or 380 day TTV signal with semi-amplitude $\sim1$~min that could be reproduced by either a weak perturber in resonance with Qatar-1b, or by a massive body in the brown dwarf regime. \citealt{mislis2015} (DM15) analyzed 12 complete Qatar-1b transits and provided updated system parameters. After reviewing all TTV data available, DM15 suggest further and more precise data to determine if TTVs exist in the system. \citet{croll2015} presented near-infrared secondary eclipse observations of Qatar-1b thermal emission showing a mid-eclipse time consistent with a circular orbit. \citealt{maciejewski2015} (GM15) analyzed 18 Qatar-1b transits to redetermine system parameters and found no evidence of periodic TTV's $\ga 1$~min. 

TTV analyses in the literature often suffer from a lack of control over the data collection and calibration processes. Since transit observations are needed on many epochs for the detection and characterization of a TTV signal, researchers typically depend on multiple observatories to collect the time-series observations. With different instrumentation and operators at each observatory, it is difficult to obtain homogeneously collected data. Lack of control over guiding precision, other sources of systematics, time synchronization, exposure durations, defocus, calibration procedures, selection of comparison star ensembles, photometric aperture radii, and detrending methods may produce veiled inaccuracies in the derived system parameters.
For these reasons, for this study we have opted to utilize observations obtained by us homogeneously with one consistent set of instruments and procedures, as we discuss in the following section.

\section{Data and Methods}\label{sec:data}

\subsection{Data from the Literature}

As part of our system analyses, we make use of radial velocity and stellar characterization data
collected from the literature. The WASP-12b analysis includes SOPHIE RV and RM data from H09 and \citet{husnoo2011}, spectroscopic parameters $\teff=6300\pm200$~K, $\feh=0.3\pm0.15$, and the nominal value of $\vsini=2200$ from H09, and orbital eccentricity $e=0$ from \citet{campo2011} and \citet{croll2011}. The Qatar-1b analysis includes TRES RV data from A11, HARPS-N RV and RM data, and spectroscopic parameters $\teff=4910\pm100$~K,  $\feh=0.2\pm0.1$, and $\vsini=1700\pm300$ m~s$^{-1}$ from C13, and orbital eccentricity $e=0$ from C13.

To maximize the precision of transit timing and other system parameters, we desire to derive all transit data homogeneously starting with consistent observational methodologies and instrumentation and following common practices through the final global fitting process. Many of the critical observational and data reduction methodologies that are discussed in \S \ref{sec:intro} cannot be controlled when analyzing transit data from the literature. Furthermore, many photometric time-series observations from the literature provide only partial transit coverage which limits the precision of the derived transit center time value. Therefore, we do not include transit data from the literature in our analyses, but include only our new homogeneously derived full transit data sets to maximize transit timing precision.

\subsection{New Observations}\label{sec:wasp12obs}

We observed 22 complete and one nearly complete high-precision transits of WASP-12b over the time span of November 2009 to February 2015 and 18 complete high-precision transits of Qatar-1b over the time span of June 2011 to September 2014. All observations were collected using the Moore Observatory Ritchie-Chretien (MORC) 0.6 m telescope. The MORC telescope was manufactured by RC Optical Systems and is operated by the University of Louisville. The instrument features a very robust fork mounting with an absolute Renishaw precision encoder on the polar axis that provides highly accurate free-running tracking of the sky. Unguided exposures of up to $\sim5$ minutes result in a tracking error of less than $1\arcs-3\arcs$, depending on the altitude of the object.

An Apogee U16M CCD camera was used to collect the time-series observations. The camera has a Kodak KAF-16803 CCD detector with a $4096\times4096$ array of $9\times9~\mu$m microlensed pixels which oversample the seeing and facilitate high precision photometry over a wide dynamic range. The pixel scale is $0\farcs39$ per pixel, which provides a field of view of $26\farcm2\times26\farcm2$. The wide field of view helps to improve differential photometry by offering a wider selection of comparison stars. The oversampled seeing improves photometry by sampling the light from a point source with more pixels (similar to telescope defocusing), improving the dynamic range of possible brightness measurements within an image, and reducing noise resulting from inter-pixel variations combined with imperfect guiding.

The date, exposure time, number of exposures, filter, photometric precision, and the error scaling factor (as determined by multi-EXOFAST; see \S\ref{sec:globalfit}) of each time-series of observations are listed in Tables \ref{tab:wasp12alltransits} and \ref{tab:qatar1alltransits} for WASP-12b and Qatar-1b, respectively. All observations were guided from the science images and were defocused to improve photometric precision. We carefully synchronized our timing source (see \S \ref{sec:timing}) and converted to $\bjdtdb$ as discussed in \citet{east10}. All exposure times were 100~s, resulting in a $\sim2$ minute cadence. All WASP-12b observations were in the Sloan $r'$ filter, except for the $g'$ observation on UT~2010-01-14, the Clear with Blue Block (CBB; a highpass filter with cutoff at $\sim 500$ nm) observation on UT~2011-12-08, and the $V$ observation on UT~2013-11-11. The $g'$ observation was the result of an error in the filter wheel controller. The CBB observation was an attempt to improve photometric precision, but it caused some of the brighter comparison stars to saturate or nearly saturate, so we reverted to $r'$ as the nominal filter for the remaining WASP-12b observations. The $V$ filter was used because the $r'$ filter was not installed in the filter wheel that night due to observations for another program. The first seven Qatar-1b transits were observed with no filter (i.e. open). After a CBB filter was acquired in late 2011, it was used to observe the remaining 11 Qatar-1b transits to reduce airmass trend in the photometric data. Multi-EXOFAST handles the observations in different filters seamlessly by extracting different limb darkening coefficients for each filter band (see \S\ref{sec:globalfit}). All transit observations include coverage of pre-ingress baseline, ingress, flat bottom, egress, and post-egress baseline, except for the WASP-12b light curve on UT~2015-01-01, which has no pre-ingress baseline or time of first contact coverage. A few of the transits have short gaps due to passing clouds or equipment problems.

\begin{table}
\begin{center}
\footnotesize
\caption{WASP-12 Photometric Observations} 
\label{tab:wasp12alltransits}
\begin{threeparttable}
{\setlength{\tabcolsep}{0.3em}
\begin{tabular}{ccccccc}
\hline
\multicolumn{1}{l}{Telescope} & \multicolumn{1}{c}{UT Date} & \multicolumn{1}{c}{Filter} & \multicolumn{1}{c}{\# Data} & \multicolumn{1}{c}{ExpT$^2$} & \multicolumn{1}{c}{RMS$^3$} & \multicolumn{1}{c}{Scale$^4$}\\
\hline
MORC$^1$ & 2009-11-05 & $r'$ & 160 & 100 & 1.3 & 1.32\\
MORC & 2009-11-28 & $r'$ & 134 & 100 & 1.0 & 1.33\\
MORC & 2010-01-13 & $r'$ & 123 & 100 & 1.1 & 1.24\\
MORC & 2010-01-14 & $g'$ & 151 & 100 & 1.2 & 1.46\\
MORC & 2010-11-09 & $r'$ & 129 & 100 & 1.2 & 1.70\\
MORC & 2010-11-10 & $r'$ & 119 & 100 & 0.9 & 1.29\\
MORC & 2011-02-11 & $r'$ & 138 & 100 & 0.9 & 1.22\\
MORC & 2011-12-08 & CBB & 148 & 100 & 0.9 & 1.30\\
MORC & 2012-02-27 & $r'$ & 167 & 100 & 1.0 & 1.31\\
MORC & 2012-02-28 & $r'$ & 135 & 100 & 1.2 & 1.43\\
MORC & 2012-03-10 & $r'$ & 148 & 100 & 1.2 & 1.40\\
MORC & 2012-11-18 & $r'$ & 135 & 100 & 1.0 & 1.26\\
MORC & 2012-12-12 & $r'$ & 139 & 100 & 1.0 & 1.09\\
MORC & 2012-12-23 & $r'$ & 180 & 100 & 1.1 & 1.27\\
MORC & 2013-01-05 & $r'$ & 160 & 100 & 1.1 & 1.20\\
MORC & 2013-01-27 & $r'$ & 194 & 100 & 1.3 & 1.07\\
MORC & 2013-11-11 & $V$ &  93 & 100 & 1.3 & 1.15\\
MORC & 2013-12-28 & $r'$ & 167 & 100 & 1.1 & 1.40\\
MORC & 2014-01-20 & $r'$ & 181 & 100 & 1.1 & 1.25\\
MORC & 2014-12-21 & $r'$ & 201 & 100 & 1.4 & 1.34\\
MORC & 2015-01-01 & $r'$ & 150 & 100 & 1.9 & 1.20\\
MORC & 2015-02-06 & $r'$ & 217 & 100 & 1.3 & 1.10\\
MORC & 2015-02-07 & $r'$ & 138 & 100 & 1.1 & 1.20\\

\hline
\end{tabular}}
\begin{tablenotes}
\item [1]MORC=U. of Louisville Moore Obs. RCOS telescope\\
\item [2]Exposure time in seconds\\
\item [3]RMS in units of $10^{-3}$\\
\item [4]Error scaling factor as determined by \multifast
\end{tablenotes}
\end{threeparttable}
\end{center}
\end{table}

\begin{table}
\begin{center}
\footnotesize
\begin{threeparttable}
\caption{Qatar-1 Photometric Observations}
\label{tab:qatar1alltransits}
{\setlength{\tabcolsep}{0.3em}
\begin{tabular}{ccccccc}
\hline
\multicolumn{1}{l}{Telescope} & \multicolumn{1}{c}{UT Date} & \multicolumn{1}{c}{Filter} & \multicolumn{1}{c}{\# Data} & \multicolumn{1}{c}{ExpT$^2$} & \multicolumn{1}{c}{RMS$^3$} & \multicolumn{1}{c}{Scale$^4$}\\
\hline
MORC$^1$ & 2011-06-30 & Open & 109 & 100 & 1.2 & 1.49\\
MORC & 2011-07-10 & Open & 120 & 100 & 1.3 & 1.47\\
MORC & 2011-08-16 & Open & 101 & 100 & 1.4 & 1.49\\
MORC & 2011-08-23 & Open & 116 & 100 & 1.1 & 1.30\\
MORC & 2011-09-22 & Open &  86 & 100 & 1.1 & 1.42\\
MORC & 2011-10-09 & Open &  83 & 100 & 1.1 & 1.22\\
MORC & 2011-12-02 & Open &  92 & 100 & 1.3 & 1.25\\
MORC & 2012-06-19 & CBB & 115 & 100 & 1.3 & 1.38\\
MORC & 2012-06-29 & CBB & 163 & 100 & 1.3 & 1.31\\
MORC & 2012-08-02 & CBB & 138 & 100 & 1.6 & 1.37\\
MORC & 2012-08-12 & CBB & 110 & 100 & 1.5 & 1.53\\
MORC & 2012-08-22 & CBB & 145 & 100 & 1.4 & 1.40\\
MORC & 2012-10-25 & CBB &  97 & 100 & 1.4 & 1.43\\
MORC & 2013-07-16 & CBB & 168 & 100 & 1.2 & 1.18\\
MORC & 2014-07-16 & CBB & 175 & 100 & 1.5 & 1.34\\
MORC & 2014-07-23 & CBB & 164 & 100 & 1.6 & 1.32\\
MORC & 2014-08-19 & CBB & 139 & 100 & 1.7 & 1.25\\
MORC & 2014-09-25 & CBB & 162 & 100 & 1.6 & 1.44\\

\hline
\end{tabular}}
\begin{tablenotes}
\item [1]MORC=U. of Louisville Moore Obs. RCOS telescope\\
\item [2]Exposure time in seconds\\
\item [3]RMS in units of $10^{-3}$\\
\item [4]Error scaling factor as determined by \multifast
\end{tablenotes}
\end{threeparttable}
\end{center}
\end{table}

\subsection{Time Stamps}\label{sec:timing}

For transit timing studies, it is critical to accurately time stamp each image with its exposure start time. Knowing the exposure start time and the exposure duration, the mid-exposure time can be calculated by data reduction software. The Linux computer that controls the MORC telescope and instrumentation synchronizes with several time servers using Network Time Protocol (NTP). The control computer communicates directly with two local tertiary stratum 2 NTP servers and takes into account the transmission delay between the server and the computer.  The tertiary servers communicate with secondary stratum 1 NTP servers around the regional network. The timing network maintains the control computer's timing accuracy to within a few microseconds of UTC1 set by the National Institute of Standards and Technology. Also, a GPS-based timing system is installed locally at the observatory and provides a stratum 1 NTP server. This server allows observations to continue if Internet connectivity is not available.

Timing errors of a few microseconds are negligible for the $>20$ second transit timing variations that we are able to detect from ground-based observations. However, due to the finite speed of light, the position of the Earth in its orbit around the Sun can cause a 15 minute peak-to-peak variation in UTC1 time from the absolute time of an extraterrestrial event. We adopt an absolute time standard based on a reference frame centered at the barycenter of the solar system, which is referred to as the Barycentric Julian Date in the Barycentric Dynamical Time ($\bjdtdb$). $\bjdtdb$ is used as the time base of all data presented in this work. \citet{east10} describe the various timing standards (including JD, UTC1, TAI, TT, etc.), the variations in terrestrial-based clocks from absolute time due to choice of reference frame, and the overall rational for the $\bjdtdb$ time base for exoplanet research.  

Our camera control software writes the UTC1 date and time of exposure start (based on the NTP synchronized Linux operating system time) and the exposure duration into the FITS header of all images. Our data reduction software package (see \S \ref{sec:wasp12dr}) reads the UTC1 date, time, and exposure duration from the FITS header and calculates $\bjdtdb$ for each image based on the coordinates of the target. The resulting mid-exposure time in $\bjdtdb$ format is used as the time-base for all plots and all global system fits presented herein.

\subsection{Data Reduction}\label{sec:wasp12dr}

We used AstroImageJ (AIJ; \citealt{collins2016}) to calibrate all images and to extract differential aperture photometry and detrending parameters from the calibrated images. AIJ has been well tested against several other commercial and open source scientific image extraction tools as part of the Kilo-degree Extremely Little Telescope (KELT; \citealt{pepper2003,pepper2007}) transit survey program, and has been found to produce photometric results that are consistent with and often more precise than results from other tools. Calibration was performed using the Data Processor module of AIJ and included bias subtraction, CCD linearity correction, dark subtraction, and flat-field division. Differential photometry was performed using the Multi-Aperture (MA) module of AIJ. Supersets of 16 and 15 comparison stars were selected for WASP-12b and Qatar-1b, respectively, ensuring that each one had brightness similar to the target star and had no nearby stars that might blend into the photometric aperture. The AIJ Multi-Plot and Light Curve Fitting modules were used to search the superset of comparison stars for a subset that minimized the best fit transit model residuals for each time-series. The final comparison ensembles were different for each time-series due to differing filters and sky transparencies, but each WASP-12b ensemble had $10-14$  comparison stars, and each Qatar-1b ensemble had $6-10$ comparison stars. A range of fixed and variable aperture sizes were investigated to find the optimal configuration that minimized transit model residuals for each dataset. The result was a variable aperture radius with a scaling factor in the range of $0.9-1.0$ times the average full-width half-maximum (FWHM) of the stellar apertures in each WASP-12b image and $1.1$ times the average FWHM in the Qatar-1b images.  

The AIJ MA module calculates photometric uncertainties, including contributions from Poisson noise in the source and sky background, CCD read-out noise, dark current, and digitization noise, and propagates the uncertainties through the differential photometry calculations. AIJ does not include a noise contribution from atmospheric scintillation. However, multi-EXOFAST scales the photometric noise to ensure the best fit individual transit model has \chisqred = 1.0 (see \S\ref{sec:globalfit}).

\begin{figure}
\begin{center}
\resizebox{\columnwidth}{!}{
\includegraphics*{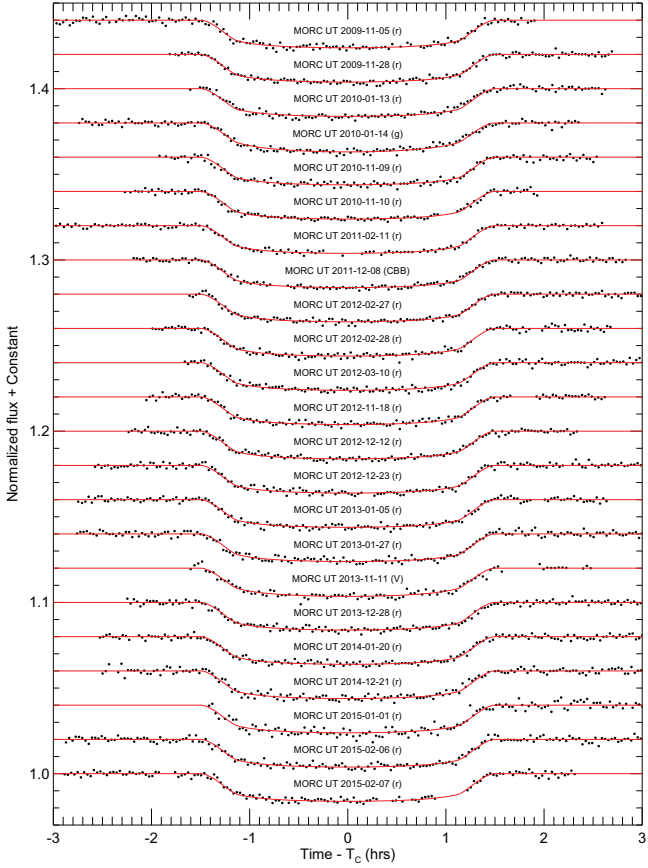} }
\caption[WASP-12b Transit Light Curves]{All 23 WASP-12b light curves. Each light curve is detrended and fitted as discussed in \S \ref{sec:wasp12globalfit}. The exposure time for each data point is 100~s. The black points are the normalized and detrended differential photometric measurements, while the red lines show the global transit model.  Each light curve is shifted on the $y$-axis for clarity.}
\label{fig:wasp12alltransits}
\end{center}
\end{figure}

\begin{figure}
\begin{center}
\resizebox{\columnwidth}{!}{
\includegraphics*{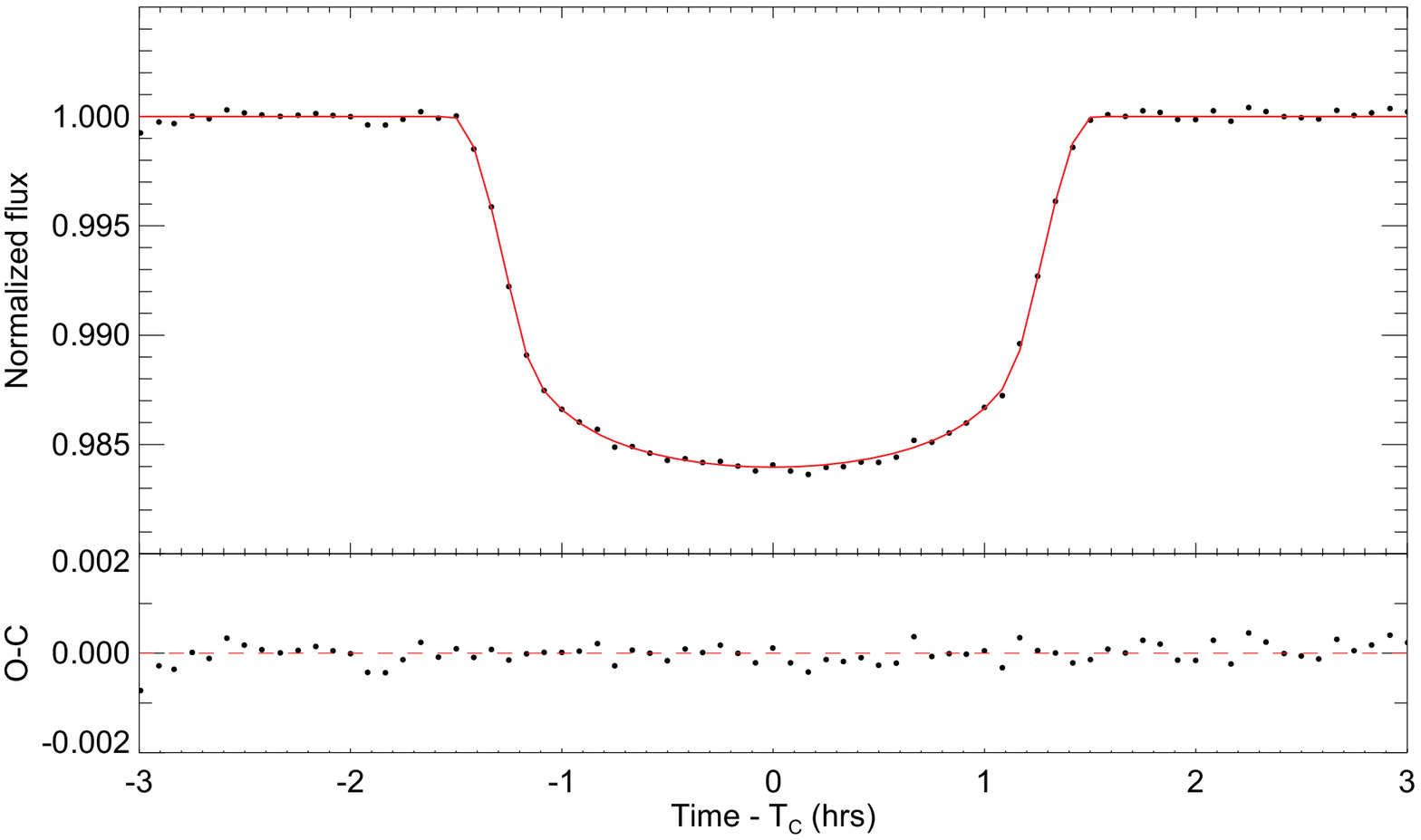} }
\caption[WASP-12b Binned Light Curve]{All 23 WASP-12b detrended light curves combined and binned in 5 min intervals (black symbols). The light curves are combined by phasing the data using the global $T_{\rm 0}$ and $P$ from Table \ref{tab:wasp12fitparmvalues}. The light curve model is also binned at 5 minute intervals and shown as a red line. This light curve is not used for analysis, but rather to show the best combined behavior of the transit. The model residuals shown in the bottom panel have an RMS of 183~ppm.
\label{fig:wasp12binnedtransits}}
\end{center}
\end{figure}

\begin{figure}
\begin{center}
\resizebox{\columnwidth}{!}{
\includegraphics*{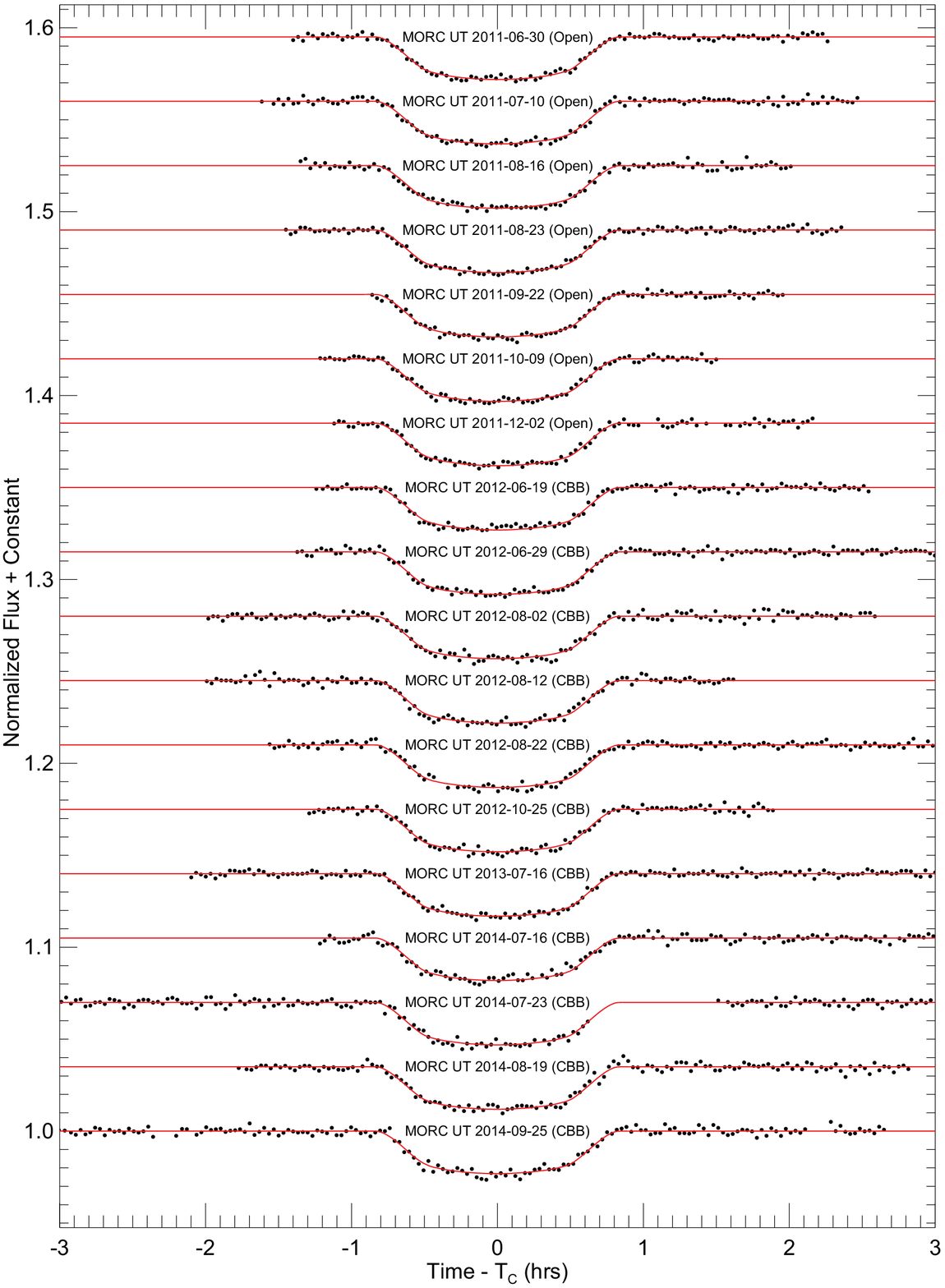} }
\caption[Qatar-1b Transit Light Curves]{All 18 Qatar-1b light curves. Each light curve is detrended and fitted as discussed in \S \ref{sec:qatar1globalfit}. The exposure time for each data point is 100~s. The black points are the normalized and detrended differential photometric measurements, while the red lines show the global transit model.  Each light curve is shifted on the $y$-axis for clarity.}
\label{fig:qatar1alltransits}
\end{center}
\end{figure}

\begin{figure}
\begin{center}
\resizebox{\columnwidth}{!}{
\includegraphics*{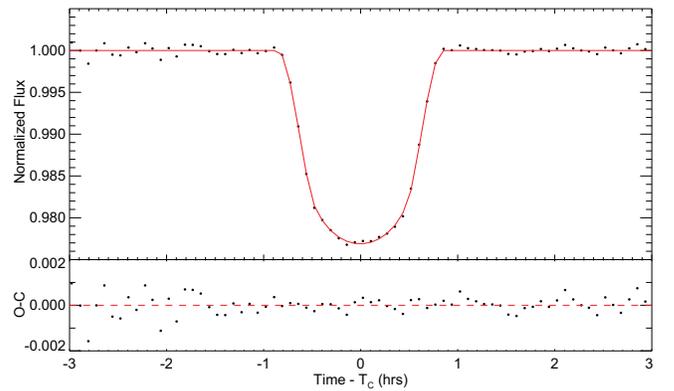} }
\caption[Qatar-1b Binned Light Curve]{All 18 Qatar-1b detrended light curves combined and binned in 5 min intervals (black symbols). The light curves are combined by phasing the data using the global $T_{\rm 0}$ and $P$ from Table \ref{tab:qatar1fitparmvalues}. The light curve model is also binned at 5 minute intervals and shown as a red line. This light curve is not used for analysis, but rather to show the best combined behavior of the transit. The model residuals are shown in the bottom panel. In the region where many transits are combined (i.e. from $-1.5$ to 2.5~hrs), the RMS of the 5 minute binned model residuals is 255~ppm. \label{fig:qatar1binnedtransits}}
\end{center}
\end{figure}

The 23 individual WASP-12b detrended and fitted light curves are shown as black symbols in Figure \ref{fig:wasp12alltransits}. The 18 Qatar-1b individual detrended and fitted light curves are shown as black symbols in Figure \ref{fig:qatar1alltransits}. The red lines show the global fit models from sections \ref{sec:wasp12globalfit} and \ref{sec:qatar1globalfit}. The black symbols in Figures \ref{fig:wasp12binnedtransits} and \ref{fig:qatar1binnedtransits} show the results after phasing (using the global fit $T_{\rm 0}$ and $P$ from Tables \ref{tab:wasp12fitparmvalues} and \ref{tab:qatar1fitparmvalues}), combining, and binning all of the light curves for each planetary system in five minute intervals. The light curve models are also combined and binned at five minute intervals and shown as red lines. The binned light curves are not used for analysis, but rather to show the best combined behavior of each set of transits. The model residuals shown in the bottom panels have an RMS of 183~ppm and 255~ppm for WASP-12b and Qatar-1b, respectively. The WASP-12b RMS value was calculated over the entire data set, while the Qatar-1b RMS value was calculated over the somewhat arbitrarily determined region in phase space from $-1.5$ to 2.5~hrs where most light curves contain data to constrain the global model. The RMS result from this limited region is a better representation of the precision of the in-transit and nearby baseline data that heavily influence the transit-derived global parameters. Each light curve is shifted on the $y$-axis for clarity.

\subsection{Global Fitting with Multi-EXOFAST}\label{sec:globalfit}

EXOFAST \citep{eastman13} provides the capability to simultaneously fit a single transit light curve, a RV data set, and host star spectroscopic parameters, and it uses Markov Chain Monte Carlo (MCMC) to robustly determine system parameter median values and uncertainties. EXOFAST uses the Torres relations \citep{torres2010} to estimate $M_*$ and $R_*$ from $\teff$, $\loggstar$, and $\feh$ at each MCMC step. Based on initial best fits to each individual data set, EXOFAST scales the RV and light curve uncertainties to target a reduced chi-squared of $\chisqred=1.0$ for each data set before starting the global fitting process. 

EXOFAST was expanded to globally model KELT transit survey planet discoveries. The resulting code, referred to as \multifast herein, provides the capability to jointly fit multiple follow-up transit light curves in differing filter bands and multiple RV telescope data sets, including the Rossiter–McLaughlin effect, and it calculates the TTV of each light curve from the linear ephemeris. \Multifast also provides the option to use the \citet{yy04} stellar models instead of the Torres relations to break the $M_*$ and $R_*$ degeneracy at each MCMC step. \Multifast treats RV data as relative velocities and includes a baseline offset fitted parameter $\gamma$ for each RV telescope as part of the global fit. The capabilities of \multifast are described in more detail in \citet{siverd2012}. Multi-EXOFAST was used to perform global fits for both the WASP-12b and Qatar-1b systems.

\citet{carter2009} show that correlated (red) noise and/or the effects of astrophysical sources such as stellar activity or variability in a light curve often lead to system parameter inaccuracies and underestimated uncertainties, if not accounted for in the model parameterization. Transit center time measurements can be particularly sensitive to light curve systematics, and to a certain extent, transit depth can be sensitive. \citet{carter2009}  describe a wavelet-based algorithm that improves the estimation of parameter values and uncertainties from light curves with correlated noise. The current version of \multifast does not implement the wavelet algorithm, but does attempt to compensate for correlated noise and underestimated input data uncertainties by scaling those uncertainties as described above for EXOFAST. Although uncertainty scaling does not perfectly account for correlated noise and/or stellar variability, it performs reasonably well when correlated systematics are not dominant. Tables \ref{tab:wasp12alltransits} and \ref{tab:qatar1alltransits} show that \multifast scales our photometric errors by small factors in the range 1.1 to 1.5, except for one WASP-12b light curve with an error scaling factor of 1.7. The low error scaling factors indicate that our calculated photometric uncertainty accounts for most of the uncertainty in our detrended photometric data, and that residual correlated noise is not dominant. Most of the photometric uncertainty scaling is likely compensating for atmospheric scintillation contributions to photometric uncertainties that are not accounted for by AIJ (see \S \ref{sec:wasp12dr}). If the uncertainty scaling technique does not fully account for stellar variability and correlated noise, our TTV scatter from a linear ephemeris may be higher than what would be expected from our quoted uncertainties. We examine our TTV scatter and uncertainties in \S \ref{sec:disc}.

\Multifast globally fits a single model parameter for each astrophysical quantity that is constrained by multiple data sets. For example, the orbital inclination must be the same for all data sets describing a transiting system. Global parameters are usually better constrained when fitted to multiple data sets, and they reduce the total number of model parameters compared to fitting all of the data sets individually. The global parameters derived from our light curves are highly constrained by 18 or more data sets, so several light curve detrending parameters can be included in the global model without over fitting the individual data sets. Based on more than five years of experience reducing MORC telescope light curves, we find that a subset of seven detrend data sets generally remove most of the time correlated (red) noise and other trends from typical MORC light curves. Those detrend data sets are airmass, time, sky background, FWHM of the average PSF in each image, total comparison star counts, and target x-centroid and y-centroid positions on the detector. For consistency, we include all seven detrend parameters along with each light curve and simultaneously detrended the light curves as part of the global fit. A separate TTV parameter was fit to each individual light curve, but the a single parameter for each of $R_P/R_*$, $a/R_*$, and $i$ was fit globally to all light curves. At each MCMC step, quadratic limb darkening parameters, $u1_\lambda$ and $u2_\lambda$, were determined for each filter band from the models of \citet{cb11} using $\teff$, $\loggstar$, and $\feh$, and the \citet{yy04} stellar models were used to break the $M_*$ and $R_*$ degeneracy. 

The publicly available \citet{cb11} limb darkening tables that are included in \multifast do not contain coefficients for the CBB or Open (i.e. the U16M CCD QE) bands. At our request, A. Claret computed them from the \citet{cb11} models. The nominal CBB and Open band coefficient values for WASP-12 and Qatar-1 are within 0.02 of the coefficients for the $R$ and \textit{Kepler} bands, respectively, which are already included in multi-EXOFAST. The difference in the coefficients is negligible for our ground-based data, so we used the $R$ and \textit{Kepler} band coefficients for the global fits. The resulting limb darkening coefficients for each filter are listed in Tables \ref{tab:wasp12fitparmvalues} and \ref{tab:qatar1fitparmvalues} for WASP-12b and Qatar-1b, respectively.

\section{Results}\label{sec:results}

\subsection{Global Fits}

\subsubsection{WASP-12b}\label{sec:wasp12globalfit}

The WASP-12b global fit included our 23 individual light curves and the SOPHIE RV data from H09 and \citet{husnoo2011} and spectroscopic priors of $\teff=6300\pm200$~K and $\feh=0.3\pm0.15$ from H09. A spectroscopic prior was not imposed on $\loggstar$, since the value derived exclusively from the light curves should be more accurate than the spectroscopic value (e.g. \citealt{mortier2013,mortier2014}). \Multifast scaled the H09 RV errors by 1.45 and the \citet{husnoo2011} RV errors by 2.59. The photometric scaling factors are listed in Table \ref{tab:wasp12alltransits} and range from 1.07 to 1.70. The orbital eccentricity was fixed to zero since secondary eclipse observations by \citet{campo2011} and \citet{croll2011} and RV observations and analysis by \citet{husnoo2011} place tight constraints on the circularity of the planet's orbit. The parameter $\vsini$ was constrained to be within 1 m~s$^{-1}$ of 2200 m~s$^{-1}$ (the nominal value from spectroscopy in H09), since the RM data do not provide a good constraint on the value. Without a tight $\vsini$ prior, the model fits would not converge. 

The $\sim1\arcs$ stellar binary discovered by \citet{bergfors2013} is blended with WASP-12 in all of the MORC photometry. The binary is four magnitudes fainter than WASP-12 in the $i'$ band, so the ratio of the blended flux to the WASP-12 flux is 1.026 in the $i'$ band, and would be slightly less in the $r'$ band used for most of the MORC observations. Using the blended flux will therefore underestimate the true transit depth and $R_P/R_*$ by $\sim2.6$\% and $\sim1.2$\%, respectively. Based on the nominal blended transit depth of $\sim14$ mmag from the literature, the non-blended depth would increase by $\sim0.35$~mmag. Any other parameters dependent on the transit depth will also be affected. Since typical $M_*$ and $R_*$ uncertainties from stellar models are $\sim10$\% (e.g. \citealt{basu2012}) and those uncertainties are not accounted for in this or other typical works, we do not attempt to account for the flux from the blended binary in the global system fit.

\begin{figure}
\begin{center}
\resizebox{\columnwidth}{!}{
\includegraphics*{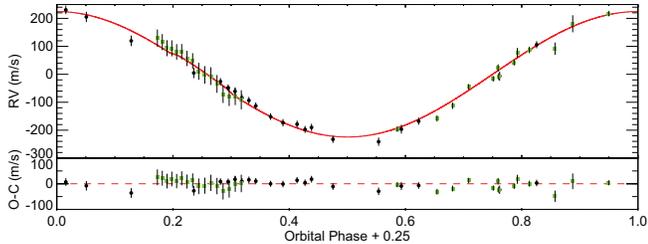} }
\caption[WASP-12b Phased RVs]{WASP-12b phased RVs and the combined RV plus RM model. The SOPHIE RVs from H09 are shown as black symbols. The SOPHIE RVs from \citet{husnoo2011} are shown as green squares. The red line shows the best fit RV plus RM model from the global fit. The bottom panel shows the model residuals.\label{fig:wasp12rvs}}
\end{center}
\end{figure}

\begin{figure}
\begin{center}
\resizebox{\columnwidth}{!}{
\includegraphics*{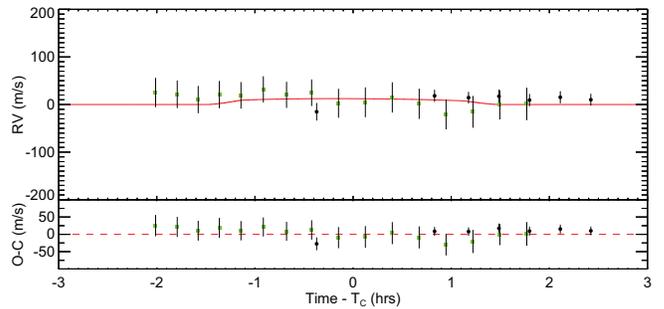} }
\caption[WASP-12b In-transit RVs]{WASP-12b phased in-transit RVs and RM model. The SOPHIE RVs from H09 are shown as black symbols. The SOPHIE RVs from \citet{husnoo2011} are shown as green squares. The red line shows the best fit RM model from the global fit. The RV data have had the best fit RV model subtracted to remove the orbital velocity component. The bottom panel shows the model residuals. \label{fig:wasp12rms}}
\end{center}
\end{figure}

The phased RVs are shown in Figure \ref{fig:wasp12rvs} and a close-up of the in-transit RM data is shown in Figure \ref{fig:wasp12rms}. The SOPHIE RVs from H09 are shown as black symbols, and the SOPHIE RVs from \citet{husnoo2011} are shown as green squares. The RM and RV models from the global fit are shown as red lines. 

\subsubsection{Qatar-1b}\label{sec:qatar1globalfit}

\begin{figure}
\begin{center}
\resizebox{\columnwidth}{!}{
\includegraphics*{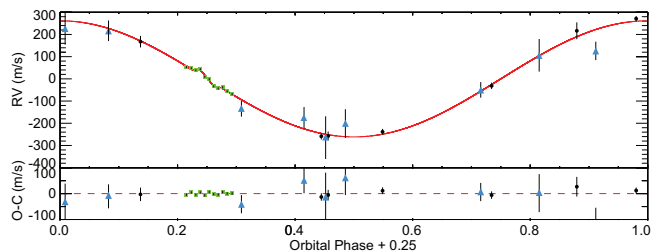} }
\caption[Qatar-1b Phased RVs]{Qatar-1b phased RVs and the combined RV plus RM model. The HARPS-N RVs and RMs from C13 are shown as circular black symbols and square green symbols, respectively. The TRES RVs from A11 are shown as blue triangles. The red line shows the best fit RV plus RM model from the global fit. The bottom panel shows the model residuals.\label{fig:qatar1rvs}}
\end{center}
\end{figure}

\begin{figure}
\begin{center}
\resizebox{\columnwidth}{!}{
\includegraphics*{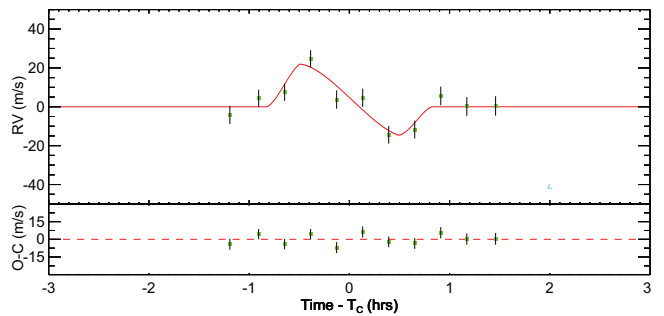} }
\caption[Qatar-1b In-transit RVs]{Qatar-1b RM data and model. The HARPS-N RMs from C13 are shown as green squares. The red line shows the best fit RM model from the global fit. The RM data have had the best fit RV model subtracted to remove the orbital velocity component. The bottom panel shows the model residuals. \label{fig:qatar1rms}}
\end{center}
\end{figure}

The Qatar-1b global fit included our 23 individual light curves, TRES RV data from A11, and HARPS-N RV and RM data, and spectroscopic priors of $\teff=4910\pm100$~K,  $\feh=0.2\pm0.1$, and $\vsini=1700\pm300$ m~s$^{-1}$ from C13. A spectroscopic prior was not imposed on $\loggstar$, as discussed in \S \ref{sec:wasp12globalfit}. \Multifast scaled the C13 RV errors by 1.93 and the A11 RV errors by 1.59. The photometric scaling factors are listed in Table \ref{tab:qatar1alltransits} and range from 1.18 to 1.53. The orbital eccentricity was fixed to zero since secondary eclipse and RV observations are consistent with a circular orbit as discussed in \S \ref{sec:intro}. The phased RVs are shown in Figure \ref{fig:qatar1rvs} and a close-up of the in-transit RM data is shown in Figure \ref{fig:qatar1rms}. The HARPS-N RVs and RMs from C13 are shown as circular black symbols and square green symbols, respectively. The TRES RVs from A11 are shown as blue triangles. The RM and RV models from the global fit are shown as red lines. 

\subsection{Improved System Parameters}

\subsubsection{WASP-12b}\label{sec:wasp12params}

In addition to the original WASP-12b system parameter analysis by H09, \citealt{chan2011,chan2012} (C12), \citet{southworth2012} (S12), and M13 have performed follow-up analyses of the WASP-12b system. The follow-up analyses use various tools to fit the transit data and determine parameter errors, and they use differing methods to constrain $M_*$ and $R_*$, but none self-consistently perform a global system analysis based on the RV data, spectroscopy, light curves, and stellar models. All of the analyses except M13 included 2 to 4 light curves. M13 simultaneously fit 19 light curves to model the system parameters, but no RV or spectroscopic parameters were included in the fit.

The system parameter values and uncertainties from a self-consistent global fit to two RV datasets, 23 MORC light curves, spectroscopic $\teff$ and $\feh$, the \citet{yy04} stellar models, and the \citet{cb11} stellar limb darkening models are shown in Table \ref{tab:wasp12fitparmvalues}, and a comparison to values from the literature is shown in Table \ref{tab:wasp12comparison}. Since M13 used a modeling tool that fits only light curves, many physical parameter values were not reported and are thus not listed in column M13 of the table. Our transit derived parameter values agree with the M13 values to well within $1\sigma$, except the highly precise orbital periods differ by $3\sigma$. This difference is discussed more in \S \ref{sec:wasp12ttvs}.   

The orbital semi-major axis, $a$, from this work agrees with the other studies within $\sim1\sigma$. All of the transit derived parameters agree within $1\sigma$ as well, except the C12 $R_P/R_*$ value is $2-3\sigma$ lower than the values listed from the other studies, which causes significant differences in some of the derived physical parameters. The C12 inclination value is higher by $1\sigma$ compared to the other studies. It is possible that one or both of the transits included in their analysis had a systematic that was not compensated for, or there could have been an astrophysical anomaly, so the C12 results are excluded from the remainder of the comparison discussion. 

The transit derived values $R_P/R_*$, $a/R_*$, and $i$ agree with the other three studies within $1\sigma$. The stellar values $M_*$, $R_*$, $\loggstar$, and $\rho_*$ all agree within $1\sigma$, although $M_*$ and $R_*$ from this work are at the upper end of the $1\sigma$ range of the other studies. The planetary parameter values $M_P$, $R_P$, and $T_{EQ}$ from this work are also higher by $\sim1\sigma$, and $\loggp$ and $\rho_P$ are smaller by $\sim1\sigma$.

The presence of a correlated non-periodic component in the SOPHIE RVs is especially obvious during the in-transit sequence (Figure \ref{fig:wasp12rms}), which was also noticed and discussed in \citet{husnoo2011}. SOPHIE is a spectrograph that is optimized for precise radial velocity measurements in the ``High Resolution'' mode. However, WASP-12b is on the faint end of the range of the 1.93~m telescope and was measured in the ``High Efficiency'' mode, which is less optimized for RV measurements. \citet{husnoo2011} discuss that the RV anomaly could be due to instrumental noise, stellar variability, or a planetary companion that is unaccounted for in the system. Given their experience with SOPHIE in ``High Efficiency'' mode, they conclude that instrumental noise is the likely cause. The impact on this study is that the value of spin-orbit angle, $\lambda=-90_{-26}^{+23}$, derived from the global fit, may not be reliable.

\begin{table*}
\centering
\footnotesize
\caption{WASP-12 System Parameter Median Values and 68\% Confidence Intervals}
\label{tab:wasp12fitparmvalues}
\begin{tabular}{lcc}
\hline
\multicolumn{1}{l}{~~~Parameter}&\multicolumn{1}{c}{Units}&\multicolumn{1}{c}{Value}\\
\hline
\multicolumn{1}{l}{Stellar Parameters:}\\
                               ~~~$M_{*}$\dotfill &Mass (\msun)\dotfill & $1.434_{-0.090}^{+0.11}$\\
                             ~~~$R_{*}$\dotfill &Radius (\rsun)\dotfill & $1.657_{-0.044}^{+0.046}$\\
                         ~~~$L_{*}$\dotfill &Luminosity (\lsun)\dotfill & $4.05_{-0.53}^{+0.54}$\\
                             ~~~$\rho_*$\dotfill &Density (cgs)\dotfill & $0.446_{-0.014}^{+0.015}$\\
                  ~~~$\log{g_*}$\dotfill &Surface gravity (cgs)\dotfill & $4.157_{-0.012}^{+0.013}$\\
                  ~~~$\teff$\dotfill &Effective temperature (K)\dotfill & $6360_{-140}^{+130}$\\
                                 ~~~$\feh$\dotfill &Metallicity\dotfill & $0.33_{-0.17}^{+0.14}$\\
           ~~~$\lambda$\dotfill &Spin-orbit alignment (degrees)\dotfill & $-90_{-26}^{+23}$\\
\multicolumn{3}{l}{Planetary Parameters:}\\
                           ~~~$a$\dotfill &Semi-major axis (AU)\dotfill & $0.02340_{-0.00050}^{+0.00056}$\\
                                 ~~~$M_{P}$\dotfill &Mass (\mj)\dotfill & $1.470_{-0.069}^{+0.076}$\\
                               ~~~$R_{P}$\dotfill &Radius (\rj)\dotfill & $1.900_{-0.055}^{+0.057}$\\
                           ~~~$\rho_{P}$\dotfill &Density (cgs)\dotfill & $0.266_{-0.014}^{+0.015}$\\
                      ~~~$\log{g_{P}}$\dotfill &Surface gravity\dotfill & $3.004\pm0.015$\\
               ~~~$T_{eq}$\dotfill &Equilibrium temperature (K)\dotfill & $2580_{-62}^{+58}$\\
                           ~~~$\Theta$\dotfill &Safronov number\dotfill & $0.02520_{-0.00084}^{+0.00087}$\\
                   ~~~$\fave$\dotfill &Incident flux (\fluxcgs)\dotfill & $10.06\pm0.94$\\
\multicolumn{3}{l}{Primary Transit Parameters:}\\
~~~$T_{\rm 0}$\dotfill & Linear ephemeris from transits (\bjdtdb)\dotfill & $2456176.668258 \pm 7.7650773\times10^{-5}$\\
~~~$P$\dotfill & Linear eph. period from transits (days)\dotfill & $1.0914203 \pm 1.4432653\times10^{-7}$\\
~~~$R_{P}/R_{*}$\dotfill &Radius of the planet in stellar radii\dotfill & $0.11785_{-0.00054}^{+0.00053}$\\
           ~~~$a/R_*$\dotfill &Semi-major axis in stellar radii\dotfill & $3.039_{-0.033}^{+0.034}$\\
                          ~~~$i$\dotfill &Inclination (degrees)\dotfill & $83.37_{-0.64}^{+0.72}$\\
                               ~~~$b$\dotfill &Impact parameter\dotfill & $0.351_{-0.034}^{+0.030}$\\
                             ~~~$\delta$\dotfill &Transit depth\dotfill & $0.01389\pm0.00013$\\
                    ~~~$T_{FWHM}$\dotfill &FWHM duration (days)\dotfill & $0.10958\pm0.00023$\\
              ~~~$\tau$\dotfill &Ingress/egress duration (days)\dotfill & $0.01526\pm0.00044$\\
                     ~~~$T_{14}$\dotfill &Total duration (days)\dotfill & $0.12483_{-0.00043}^{+0.00044}$\\
   ~~~$P_{T}$\dotfill &A priori non-grazing transit probability\dotfill & $0.2903_{-0.0031}^{+0.0030}$\\
              ~~~$P_{T,G}$\dotfill &A priori transit probability\dotfill & $0.3679_{-0.0042}^{+0.0041}$\\
\multicolumn{3}{l}{Secondary Eclipse Parameter:}\\
                  ~~~$T_{S}$\dotfill &Time of eclipse (\bjdtdb)\dotfill & $2456176.12256\pm0.00020$\\
\multicolumn{3}{l}{RV Parameters:}\\
                        ~~~$K$\dotfill &RV semi-amplitude (m/s)\dotfill & $226.4\pm4.1$\\
                      ~~~$e$\dotfill &RV eccentricity\dotfill & $0.0$ (FIXED)\\
                           ~~~$K_{\rm RM}$\dotfill &RM amplitude (m/s)\dotfill & $30.99_{-0.29}^{+0.28}$\\
                    ~~~$M_P\sin{i}$\dotfill &Minimum mass (\mj)\dotfill & $1.460_{-0.068}^{+0.075}$\\
                           ~~~$M_{P}/M_{*}$\dotfill &Mass ratio\dotfill & $0.000977\pm0.000028$\\
                            ~~~$\gamma_{SOPHIE09}$\dotfill &m/s\dotfill & $72.4\pm3.6$\\
                            ~~~$\gamma_{SOPHIE10}$\dotfill &m/s\dotfill & $19083.0\pm3.1$\\
\multicolumn{3}{l}{Limb-darkening Coefficients:}\\
~~~$u_{1Kepler~(for~Open)}$\dotfill &Linear Limb-darkening\dotfill & $0.322_{-0.017}^{+0.018}$\\
             ~~~$u_{2Kepler~(for~Open)}$\dotfill &Quadratic Limb-darkening\dotfill & $0.3124_{-0.0098}^{+0.0091}$\\
                ~~~$u_{1Sloang}$\dotfill &Linear Limb-darkening\dotfill & $0.481_{-0.023}^{+0.024}$\\
             ~~~$u_{2Sloang}$\dotfill &Quadratic Limb-darkening\dotfill & $0.271_{-0.015}^{+0.014}$\\
                ~~~$u_{1Sloanr}$\dotfill &Linear Limb-darkening\dotfill & $0.312_{-0.017}^{+0.018}$\\
             ~~~$u_{2Sloanr}$\dotfill &Quadratic Limb-darkening\dotfill & $0.3264_{-0.0097}^{+0.0088}$\\
                     ~~~$u_{1V}$\dotfill &Linear Limb-darkening\dotfill & $0.382_{-0.018}^{+0.019}$\\
                  ~~~$u_{2V}$\dotfill &Quadratic Limb-darkening\dotfill & $0.3080_{-0.0100}^{+0.0096}$\\
\hline
\end{tabular}
\vspace{6 pt}
\end{table*}

\begin{table*}
\begin{center}
\footnotesize
\begin{threeparttable}
\caption{Comparison of WASP-12 System Parameters to Literature Values}
\label{tab:wasp12comparison}
\begin{tabular}{cccccc}
\hline
\multicolumn{1}{l}{Parameter} & \multicolumn{1}{c}{This Work} & \multicolumn{1}{c}{M13} & \multicolumn{1}{c}{S12} & \multicolumn{1}{c}{C12} & \multicolumn{1}{c}{H09}\\
\hline
\multicolumn{3}{l}{Stellar Parameters:}\\
$M_*$        & $1.434\pm0.11$         & $--$                   & $1.38\pm0.18$       & $1.36\pm0.14$ & $1.35\pm0.14$\\
$R_*$        &  $1.657\pm0.046$       & $--$                   & $1.619\pm0.076$   & $1.595\pm0.071$ & $1.57\pm0.07$\\
$\loggstar$ & $4.157\pm0.013$        & $--$                   & $4.159\pm0.023$   & $4.164\pm0.029$ & $4.17\pm0.03$\\
$\rho_*$    & $0.446\pm0.015$        & $0.444\pm0.01$ & $0.458\pm0.023$  & $0.475\pm0.038$ & $0.493\pm0.04$\\
\multicolumn{3}{l}{Planetary Parameters:}\\
$M_P$       & $1.470\pm0.076$         & $--$                 & $1.43\pm0.13$        & $1.403\pm0.099$ & $1.41\pm0.10$\\
$R_P$        & $1.900\pm0.057$        & $1.86\pm0.09$  & $1.825\pm0.091$    & $1.732\pm0.092$ & $1.79\pm0.09$\\
$\loggp$    & $3.004\pm0.015$        & $--$                  & $3.027\pm0.023$    & $3.069\pm0.031$  & $2.99\pm0.03$\\
$\rho_P$    & $0.266\pm0.015$        & $--$                 & $0.292\pm0.025$     & $0.340\pm0.039$ & $0.318\pm0.04$\\
$T_{eq}$   & $2580\pm62$              & $--$                & $2523\pm45$           & $--$                    & $2516\pm36$\\
\multicolumn{3}{l}{Orbital Parameters:}\\
$P$           & $1.0914203(1)$           & $1.0914209(2)$  & $--$                      & $1.0914222(11)$ & $1.0914230(30)$\\
$a$           & $0.02340\pm0.00056$  & $--$                & $0.02309\pm0.00096$ & $0.02293\pm0.00078$ & $0.0229\pm0.0008$\\
\multicolumn{3}{l}{Transit Parameters:}\\
$R_P/R_*$ & $0.11785\pm0.00054$  & $0.1173\pm0.0005$ & $0.1158\pm0.005$ & $0.1119\pm0.002$ & $0.1175\pm0.0008$\\
$a/R_*$    & $3.039\pm0.034$         & $3.033\pm0.022$ & $3.067\pm0.05$    & $3.105\pm0.082$   & $--$\\
$i$            & $83.37\pm0.72$          & $82.96\pm0.50$  & $83.3\pm1.1$         & $86.2\pm3.0$       & $83.1\pm1.4$\\
\hline
\end{tabular}
\begin{tablenotes}
\item [Notes:]Parameter variable names and units are the same as in Table \ref{tab:wasp12fitparmvalues}, a value enclosed in parentheses is the uncertainty in the same number of last digits, M13=\citet{maciejewski2013}, S12=\citet{southworth2012}, C12=\citet{chan2011,chan2012}, H09=\citet{hebb2009}
\end{tablenotes}
\end{threeparttable}
\end{center}
\vspace{6 pt}
\end{table*}

\subsubsection{Qatar-1b}\label{sec:qatar1params}

In addition to the original Qatar-1b system analysis by A11, multi-light curve analyses have been performed by C13, E13, DM15, and GM15. The follow-up analyses use various tools to fit the transit data and determine parameter errors, and they use differing methods to constrain $M_*$ and $R_*$, but none self-consistently perform a global system analysis based on the RV data, spectroscopy, light curves, and stellar models. 

The system parameter values and uncertainties from a self-consistent global fit to two RV datasets, 18 MORC light curves, spectroscopic $\teff$, $\feh$, and $\vsini$, the \citet{yy04} stellar models, and the \citet{cb11} stellar limb darkening models are shown in Table \ref{tab:qatar1fitparmvalues}, and a comparison to values from the literature is shown in Table \ref{tab:qatar1comparison}. Since E13 only reported values determined exclusively from the transit model fit, physical parameter values are not listed in column E13 of the table. 

The stellar parameters, $M_*$, $R_*$, $\loggstar$, and $\rho_*$ from this work agree with values from all five literature sources to well within $1\sigma$. C13 revised the RV semi-amplitude upward based on new HARPS-N velocities, which resulted in an increase in planet mass. Since we include the new RVs in this global fit, our planetary parameters are best compared to the literature values starting with C13. They all agree to within $1\sigma$ and are almost identical to the DM15 results, except our $T_{eq}$ is higher by about $1\sigma$. The orbital and transit parameters agree within $1\sigma$, except $R_P/R_*$ from C13 and E13 differ from all others by more than $3\sigma$ and $i$ from E13 is high by $2\sigma$.

In summary, the parameters from this work agree well with the literature values, except for the few outliers mentioned in the previous paragraph.

\begin{table*}
\centering
\footnotesize
\caption{Qatar-1 System Parameter Median Values and 68\% Confidence Intervals}
\label{tab:qatar1fitparmvalues}
\begin{tabular}{lcc}
\hline
\multicolumn{1}{l}{~~~Parameter}&\multicolumn{1}{c}{Units}&\multicolumn{1}{c}{Value}\\
\hline
\multicolumn{1}{l}{Stellar Parameters:}\\
                               ~~~$M_{*}$\dotfill &Mass (\msun)\dotfill & $0.838_{-0.041}^{+0.043}$\\
                             ~~~$R_{*}$\dotfill &Radius (\rsun)\dotfill & $0.803\pm0.016$\\
                         ~~~$L_{*}$\dotfill &Luminosity (\lsun)\dotfill & $0.365_{-0.034}^{+0.040}$\\
                             ~~~$\rho_*$\dotfill &Density (cgs)\dotfill & $2.286_{-0.070}^{+0.074}$\\
                  ~~~$\log{g_*}$\dotfill &Surface gravity (cgs)\dotfill & $4.552_{-0.011}^{+0.012}$\\
                  ~~~$\teff$\dotfill &Effective temperature (K)\dotfill & $5013_{-88}^{+93}$\\
                                 ~~~$\feh$\dotfill &Metallicity\dotfill & $0.171_{-0.094}^{+0.097}$\\
             ~~~$v\sin{I_*}$\dotfill &Rotational velocity (m/s)\dotfill & $1760\pm210$\\
           ~~~$\lambda$\dotfill &Spin-orbit alignment (degrees)\dotfill & $-7.5_{-7.6}^{+7.5}$\\
\multicolumn{3}{l}{Planetary Parameters:}\\
                           ~~~$a$\dotfill &Semi-major axis (AU)\dotfill & $0.02332_{-0.00038}^{+0.00040}$\\
                                 ~~~$M_{P}$\dotfill &Mass (\mj)\dotfill & $1.294_{-0.049}^{+0.052}$\\
                               ~~~$R_{P}$\dotfill &Radius (\rj)\dotfill & $1.143_{-0.025}^{+0.026}$\\
                           ~~~$\rho_{P}$\dotfill &Density (cgs)\dotfill & $1.076_{-0.053}^{+0.057}$\\
                      ~~~$\log{g_{P}}$\dotfill &Surface gravity\dotfill & $3.390\pm0.015$\\
               ~~~$T_{eq}$\dotfill &Equilibrium temperature (K)\dotfill & $1418_{-27}^{+28}$\\
                           ~~~$\Theta$\dotfill &Safronov number\dotfill & $0.0629\pm0.0019$\\
                   ~~~$\fave$\dotfill &Incident flux (\fluxcgs)\dotfill & $0.918_{-0.069}^{+0.075}$\\
\multicolumn{3}{l}{Primary Transit Parameters:}\\
~~~$T_{\rm 0}$\dotfill & Linear ephemeris from transits (\bjdtdb)\dotfill & $2456234.103218\pm6.0708415\times10^{-5}$\\
~~~$P$\dotfill & Linear eph. period from transits (days)\dotfill & $1.4200242\pm2.1728848\times10^{-7}$\\
~~~$R_{P}/R_{*}$\dotfill &Radius of the planet in stellar radii\dotfill & $0.14629_{-0.00064}^{+0.00063}$\\
           ~~~$a/R_*$\dotfill &Semi-major axis in stellar radii\dotfill & $6.247_{-0.065}^{+0.067}$\\
                          ~~~$i$\dotfill &Inclination (degrees)\dotfill & $84.08_{-0.15}^{+0.16}$\\
                               ~~~$b$\dotfill &Impact parameter\dotfill & $0.645_{-0.011}^{+0.010}$\\
                             ~~~$\delta$\dotfill &Transit depth\dotfill & $0.02140\pm0.00019$\\
                    ~~~$T_{FWHM}$\dotfill &FWHM duration (days)\dotfill & $0.05498\pm0.00020$\\
              ~~~$\tau$\dotfill &Ingress/egress duration (days)\dotfill & $0.01423_{-0.00038}^{+0.00039}$\\
                     ~~~$T_{14}$\dotfill &Total duration (days)\dotfill & $0.06921\pm0.00033$\\
   ~~~$P_{T}$\dotfill &A priori non-grazing transit probability\dotfill & $0.1367_{-0.0014}^{+0.0013}$\\
              ~~~$P_{T,G}$\dotfill &A priori transit probability\dotfill & $0.1835\pm0.0020$\\
\multicolumn{3}{l}{Secondary Eclipse Parameter:}\\
                  ~~~$T_{S}$\dotfill &Time of eclipse (\bjdtdb)\dotfill & $2456233.3931\pm0.0021$\\
\multicolumn{3}{l}{RV Parameters:}\\
                        ~~~$K$\dotfill &RV semi-amplitude (m/s)\dotfill & $261.6\pm5.5$\\
                                ~~~$e$\dotfill &RV eccentricity\dotfill & $0.0$ (FIXED)\\
                           ~~~$K_{\rm RM}$\dotfill &RM amplitude (m/s)\dotfill & $38.5\pm4.6$\\
                    ~~~$M_P\sin{i}$\dotfill &Minimum mass (\mj)\dotfill & $1.287_{-0.048}^{+0.052}$\\
                           ~~~$M_{P}/M_{*}$\dotfill &Mass ratio\dotfill & $0.001474\pm0.000039$\\
                               ~~~$\gamma_{HARPS}$\dotfill &m/s\dotfill & $-50.5_{-6.1}^{+6.0}$\\
                             ~~~$\gamma_{HARPS_{rm}}$\dotfill &m/s\dotfill & $-59.6\pm2.0$\\
                                ~~~$\gamma_{TRES}$\dotfill &m/s\dotfill & $118\pm16$\\
\multicolumn{3}{l}{Limb-darkening Coefficients:}\\                                
~~~$u_{1Kepler~(for~Open)}$\dotfill &Linear Limb-darkening\dotfill & $0.577_{-0.022}^{+0.021}$\\
             ~~~$u_{2Kepler~(for~Open)}$\dotfill &Quadratic Limb-darkening\dotfill & $0.148\pm0.016$\\
                     ~~~$u_{1R~(for~CBB)}$\dotfill &Linear Limb-darkening\dotfill & $0.534_{-0.021}^{+0.020}$\\
                  ~~~$u_{2R~(for~CBB)}$\dotfill &Quadratic Limb-darkening\dotfill & $0.177\pm0.014$\\
\hline
\end{tabular}
\vspace{6 pt}
\end{table*}

\begin{table*}
\begin{center}
\footnotesize
\begin{threeparttable}
\caption{Comparison of Qatar-1 System Parameters to Literature Values}
\label{tab:qatar1comparison}
\begin{tabular}{ccccccc}
\hline
\multicolumn{1}{l}{Parameter} & \multicolumn{1}{c}{This Work} & \multicolumn{1}{c}{DM15} & \multicolumn{1}{c}{GM15} & \multicolumn{1}{c}{C13} & \multicolumn{1}{c}{E13} & \multicolumn{1}{c}{A11}\\
\hline
\multicolumn{3}{l}{Stellar Parameters:}\\
$M_*$        & $0.838\pm0.043$       & $0.818\pm0.047$    & $0.803\pm0.072$       & $0.850\pm0.030$     & $--$                      & $0.850\pm0.030$\\
$R_*$        &  $0.803\pm0.016$      & $0.796\pm0.016$     & $0.782\pm0.025$       & $0.800\pm0.050$    & $--$                       & $0.823\pm0.025$\\
$\loggstar$ & $4.552\pm0.012$       & $4.549\pm0.011$     & $4.556\pm0.020$       & $4.550\pm0.100$    & $--$                       & $4.536\pm0.024$\\
$\rho_*$    & $2.286\pm0.074$       & $2.282\pm0.065$     & $2.365\pm0.079$       & $2.281\pm0.113$    & $--$                       & $2.140\pm0.169$\\
\multicolumn{3}{l}{Planetary Parameters:}\\
$M_P$       & $1.294\pm0.052$       & $1.293\pm0.052$      & $1.275\pm0.079$       & $1.330\pm0.050$    & $--$                       & $1.090\pm0.084$\\
$R_P$        & $1.143\pm0.026$      & $1.142\pm0.026$       & $1.136\pm0.037$       & $1.180\pm0.090$    & $--$                       & $1.164\pm0.045$\\
$\loggp$    & $3.390\pm0.015$      & $3.390\pm0.012$       & $3.409\pm0.012$       & $3.372\pm0.024$    & $--$                        & $3.265\pm0.045$\\
$\rho_P$    & $1.076\pm0.057$      & $1.075\pm0.048$       & $1.154\pm0.133$        & $1.061\pm0.265$    & $--$                       & $0.915\pm0.130$\\
$T_{eq}$   & $1418\pm28$            & $1388\pm29$             & $--$                          & $1389\pm39$        & $--$                        & $1399\pm42$\\
\multicolumn{3}{l}{Orbital Parameters:}\\
$P$           & $1.4200242(2)$         & $1.4200259(28)$           & $1.4200241(2)$           & $1.4200250(7)$      & $1.4200246(4)$       & $1.4200330(160)$\\
$a$          & $0.02332\pm0.00040$ & $0.02313\pm0.00044$ & $0.02298\pm0.00069$  & $0.02343\pm0.00120$ & $--$                   & $0.02343\pm0.00026$\\
\multicolumn{3}{l}{Transit Parameters:}\\
$R_P/R_*$ & $0.1463\pm0.0006$  & $0.1475\pm0.0009$     & $0.1459\pm0.0008$      & $0.1513\pm0.0008$ & $0.1435\pm0.0008$ & $0.1455\pm0.0016$\\
$a/R_*$    & $6.25\pm0.07$         & $6.25\pm0.08$            & $6.32\pm0.07$             & $6.25\pm0.10$      & $6.42\pm0.10$       & $--$\\
$i$            & $84.08\pm0.16$       & $84.03\pm0.16$          & $84.26\pm0.17$           & $83.82\pm0.25$      & $84.52\pm0.24$      & $83.47\pm0.40$\\
\hline
\end{tabular}
\begin{tablenotes}
\item [Notes:]Parameter variable names and units are the same as in Table \ref{tab:qatar1fitparmvalues}, a value enclosed in parentheses is the uncertainty in the same number of last digits, DM15=\citet{mislis2015}, GM15=\citet{maciejewski2015}, C13=\citet{covino2013}, E13=\citet{essen2013}, A11=\citet{al11}
\end{tablenotes}
\end{threeparttable}
\end{center}
\vspace{6 pt}
\end{table*}

\subsection{Transit Timing Variation Measurements}

\subsubsection{WASP-12b}\label{sec:wasp12ttvs}

\begin{figure}
\begin{center}
\resizebox{\columnwidth}{!}{
\includegraphics*{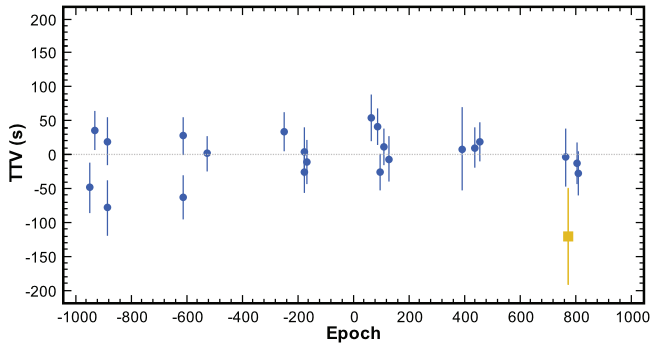} }
\caption[WASP-12b TTVs]{WASP-12b TTVs vs. transit epoch. The epoch is calculated from the global fit ephemeris in Table \ref{tab:wasp12fitparmvalues}. TTV is defined as the observed $T_{C}$ minus $T_{C}$ calculated from the linear ephemeris (i.e. $O-C$). The epoch 776 TTV with large uncertainty and shown as a square yellow symbol is from the transit on UT~2015-01-01 that does not have coverage of pre-ingress baseline or time of first contact and has been excluded from the analysis. \label{fig:wasp12ttvslinear}}
\end{center}
\end{figure}

The WASP-12b global system model includes a TTV parameter for each of the 23 light curves, allowing the transit center time at each epoch to differ from the linear ephemeris. The resulting TTV values are listed in Table \ref{tab:wasp12ttvs} and a linear plot of TTV vs. epoch is shown in Figure \ref{fig:wasp12ttvslinear}. The epoch 776 TTV with large uncertainty and shown as a square yellow symbol is from the transit on UT~2015-01-01 which does not have coverage of pre-ingress baseline or time of first contact and has been excluded from the analysis. This data point illustrates the importance of including only complete transits with good baseline in high precision TTV studies. Ignoring epoch 776, the TTVs have a maximum value of 79~s, a standard deviation of 32.88~s, and reduced chi-squared of $\chisqred=1.1$, with respect to the linear ephemeris. The mean of the timing uncertainty is 31~s.

\begin{table}
\centering
\begin{threeparttable}
{\setlength{\tabcolsep}{0.3em}
\caption{WASP-12b Transit Times}
\label{tab:wasp12ttvs}
\begin{tabular}{lrrrr}
\hline
\multicolumn{1}{c}{$T_C~(\bjdtdb)$} & \multicolumn{1}{c}{Epoch} & \multicolumn{1}{c}{TTV (s)} & \multicolumn{1}{c}{$\sigma_{\rm{TTV}}$ (s)} & \multicolumn{1}{c}{TTV/$\sigma_{\rm{TTV}}$}\\
\hline
 2455140.909815 & -949 &  -48.89 & 36 &  -1.33\\
 2455163.830613 & -928 &   35.05 & 28 &   1.21\\
 2455209.668946 & -886 &  -79.01 & 40 &  -1.93\\
 2455210.761506 & -885 &   19.46 & 35 &   0.55\\
 2455509.809713 & -611 &  -63.37 & 32 &  -1.95\\
 2455510.902181 & -610 &   27.15 & 27 &   0.97\\
 2455603.672606 & -525 &    1.09 & 25 &   0.04\\
 2455903.813566 & -250 &   33.41 & 28 &   1.19\\
 2455984.577971 & -176 &  -26.91 & 28 &  -0.93\\
 2455985.669747 & -175 &    3.82 & 36 &   0.10\\
 2455996.583783 & -165 &  -10.62 & 32 &  -0.32\\
 2456249.794041 &   67 &   53.79 & 34 &   1.58\\
 2456273.805140 &   89 &   41.01 & 26 &   1.57\\
 2456284.718565 &   99 &  -26.22 & 26 &  -1.00\\
 2456297.816050 &  111 &   11.90 & 26 &   0.45\\
 2456319.644239 &  131 &   -6.87 & 33 &  -0.21\\
 2456607.779376 &  395 &    8.21 & 61 &   0.13\\
 2456654.710469 &  438 &    9.90 & 29 &   0.33\\
 2456677.630387 &  459 &   17.80 & 28 &   0.62\\
 2457012.696166 &  766 &   -4.41 & 42 &  -0.10\\
 2457023.609018* &  776 & -121.14 & 71 &  -1.69\\
 2457059.627132 &  809 &  -13.69 & 30 &  -0.45\\
 2457060.718388 &  810 &  -27.89 & 31 &  -0.89\\
\hline
Standard Deviation &     &   32.88 & $--$ &  1.00\\ 
Mean                    &      &    $--$ & 31    & $--$\\ 
\hline     
\end{tabular}}
\begin{tablenotes}
\item* This measurement is from the transit on UT~2015-01-01 that does not have coverage of pre-ingress baseline or time of first contact and has been excluded from the analysis.
\end{tablenotes}
\end{threeparttable}
\end{table}

\subsubsection{Qatar-1b}\label{sec:qatar1ttvs}

The Qatar-1b global system model includes a TTV parameter for each of the 18 light curves, allowing the transit center time at each epoch to differ from the linear ephemeris. The resulting TTV values are listed in Table \ref{tab:qatar1ttvs} and a linear plot of TTV vs. epoch is shown in the top panel of Figure \ref{fig:qatar1ttvslinear}. The TTVs have a maximum absolute value of 47~s, RMS$=23.32$~s, and $\chisqred=1.08$, with respect to the linear ephemeris. The mean of the timing uncertainty is 22~s.

\begin{table}
\centering
{\setlength{\tabcolsep}{0.3em}
\caption{Qatar-1b Transit Times}
\label{tab:qatar1ttvs}
\begin{tabular}{lrrrr}
\hline
\multicolumn{1}{c}{$T_C~(\bjdtdb)$} & \multicolumn{1}{c}{Epoch} & \multicolumn{1}{c}{TTV (s)} & \multicolumn{1}{c}{$\sigma_{\rm{TTV}}$ (s)} & \multicolumn{1}{c}{TTV/$\sigma_{\rm{TTV}}$}\\
\hline
 2455742.774748 & -346 &  -7.41 &  19 &  -0.38\\
 2455752.714992 & -339 &  -0.98 &  21 &  -0.05\\
 2455789.635401 & -313 & -20.08 &  22 &  -0.88\\
 2455796.735833 & -308 &   6.78 &  18 &   0.36\\
 2455826.556175 & -287 &  -7.64 &  19 &  -0.39\\
 2455843.596639 & -275 &   7.33 &  19 &   0.37\\
 2455897.557680 & -237 &  17.71 &  21 &   0.83\\
 2456097.780697 &  -96 & -16.82 &  20 &  -0.83\\
 2456107.721518 &  -89 &  39.46 &  18 &   2.09\\
 2456141.801755 &  -65 &   9.69 &  25 &   0.38\\
 2456151.741312 &  -58 & -43.24 &  24 &  -1.74\\
 2456161.681666 &  -51 & -27.31 &  23 &  -1.17\\
 2456225.583358 &   -6 &  24.67 &  23 &   1.07\\
 2456489.707616 &  180 &   3.14 &  17 &   0.18\\
 2456854.653255 &  437 & -47.71 &  27 &  -1.70\\
 2456861.754168 &  442 &  20.71 &  28 &   0.74\\
 2456888.734234 &  461 & -13.37 &  26 &  -0.51\\
 2456925.655231 &  487 &  18.34 &  22 &   0.81\\
\hline
Standard Deviation &     &   23.32 & $--$ &  1.00\\ 
Mean                    &      &    $--$ & 22    & $--$\\ 
\hline     
\end{tabular}}
\end{table}

\section{Discussion and Comparison to Previous TTV Results}\label{sec:disc}

\begin{figure}
\begin{center}
\resizebox{\columnwidth}{!}{
\includegraphics*{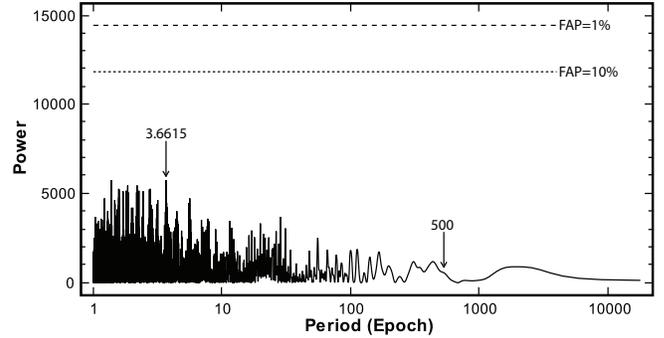} }
\caption[WASP-12b TTV Periodogram]{Lomb-Scargle periodogram for the WASP-12b TTVs. Analytical FAP levels of 10\% and 1\% are indicated by the short-dashed and long-dashed lines, respectively. The down-pointing arrow labeled 3.6615 marks the peak power component. The 500 epoch signal that was detected and investigated by M13 is marked with
the down-pointing arrow labeled 500 and shows very little power in our data. \label{fig:wasp12periodogram}}
\end{center}
\end{figure}

\begin{figure}
\begin{center}
\resizebox{\columnwidth}{!}{
\includegraphics*{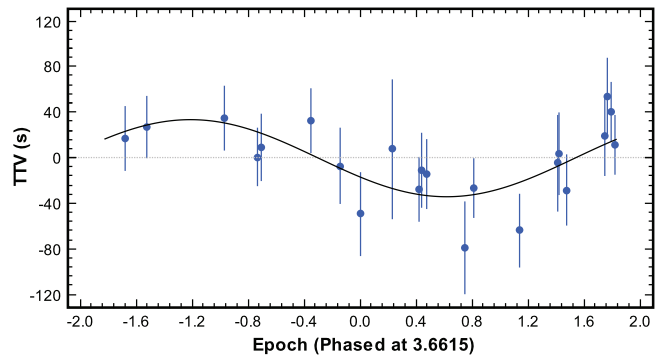} }
\caption[WASP-12b TTVs Phased at 4.0 d]{WASP-12b TTVs phased at 3.6615 epochs (3.996 days). The sinusoidal fit has $\chisqred=0.66$ and a semi-amplitude of 34~s. The interpretation is that the TTV signal is the result of a night-to-night systematic or a chance fit to noise (see text).\label{fig:wasp12ttvsphased3p6615e}}
\end{center}
\end{figure}

Figure \ref{fig:wasp12periodogram} shows a Lomb-Scargle periodogram of our WASP-12b TTVs. All periodograms and associated false alarm probabilities in this work were calculated using the Systemic Console package, as described in \citet{meschiari2009}, by loading the epoch values as the x-axis data set and the TTV values and uncertainties as the y-axis data set. All peaks are well below the analytical 10\% false alarm probably (FAP) indicated by the short-dashed line. Nevertheless, we investigated the highest power peak at 3.6615 epochs (3.996 days), which is marked in Figure \ref{fig:wasp12periodogram} with a down-pointing arrow labeled 3.6615. The phased plot and model are shown in Figure \ref{fig:wasp12ttvsphased3p6615e}. The 3.996 day period is close to 4.0 days, which suggests a systematic related to the Earth's rotation, but we cannot explain a specific possible cause. The sinusoidal fit has an improved $\chisqred=0.66$ compared to the linear fit. Based on the difference in the two $\chisq$ values and a difference of two degrees of freedom between the linear and sinusoidal fits, $\Delta\chisq$ analysis suggests a 0.6\% probability of a chance improvement in the sinusoidal fit. However, the $\sim 4.0$ day peak has an analytical false alarm probability of more than 100\%. The semi-amplitude of the TTV signal is 34~s, which is nearly the same as the standard deviation of the TTVs and the mean timing uncertainty. Our interpretation is that the TTV signal is the result of a night-to-night systematic or a chance fit to noise. The 500 epoch signal that was detected and investigated by M13 is marked in Figure \ref{fig:wasp12periodogram} with the down-pointing arrow labeled 500 and shows very little power in the periodogram calculated from our data. Time domain searches around 500 epochs confirm the results of our periodogram.

\begin{figure}
\begin{center}
\resizebox{\columnwidth}{!}{
\includegraphics*{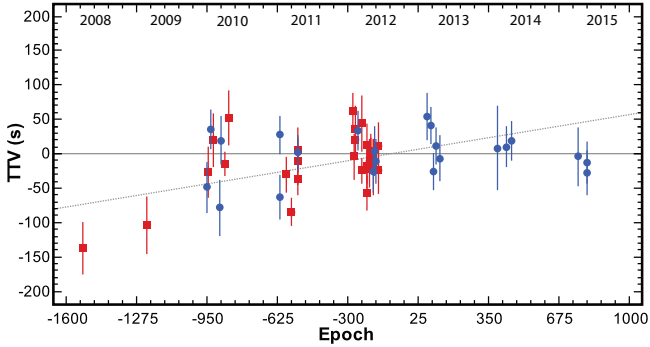} }
\caption[WASP-12b TTVs with Literature TTVs]{WASP-12b TTVs from this work and the literature. TTVs from this work are shown as circular blue symbols. TTVs collected by M13 with timing errors $<40$~s have been rephased to the ephemeris from this work and are shown as square red symbols. The solid gray line shows the linear ephemeris from this work. The dashed gray line shows the linear ephemeris from M13. The numbers across the top indicate the approximate calender year.\label{fig:wasp12literaturettvslinear}}
\end{center}
\end{figure}

M13 included TTVs from an early follow-up light curve from the H09 discovery paper, and two more early light curves from C12. Figure \ref{fig:wasp12literaturettvslinear} shows the TTVs analyzed by M13 with timing errors $<40$~s (square red symbols), after rephasing them to the linear ephemeris derived in this work (solid gray line). The TTVs from this work are indicated by circular blue symbols. The transits at epochs $-1522$ and $-1224$ arrive significantly early relative to the ephemeris from this work, but not quite so early relative to the M13 linear ephemeris (dashed gray line).

Combining all of the TTV data, a fit to the linear ephemerides gives $\chisqred=1.83$ (ephemeris from this work) and $\chisqred=1.71$ (ephemeris from M13). Excluding the early epochs $-1522$ and $-1224$ gives $\chisqred=1.48$ (this work) and $\chisqred=1.69$ (M13). An even longer time baseline of observations may be required to determine which ephemeris best describes the WASP-12b orbital period. 

The periodogram of the combined TTVs was also investigated. Again, there were no peaks with a reasonably low FAP, and in particular, little support was found for the 500 epoch signal in either the frequency or time domain.

\begin{figure}
\begin{center}
\resizebox{\columnwidth}{!}{
\includegraphics*{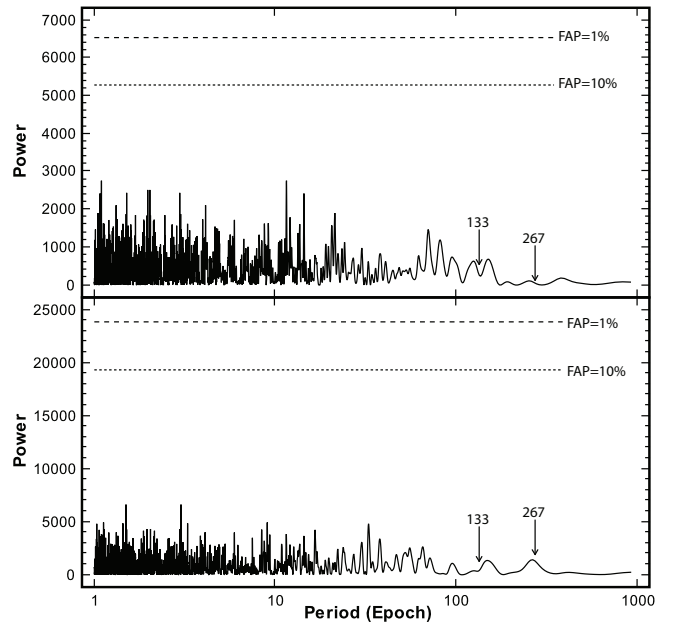} }
\caption[Qatar-1b TTV Periodograms]{Lomb-Scargle periodograms for the Qatar-1b TTVs. Analytical FAP levels of 10\% and 1\% are indicated by the short-dashed and long-dashed lines, respectively. The down-pointing arrows labeled 133 and 267 epochs mark the $\sim190$ and $\sim380$~d periods investigated by E13. \textit{Top Panel:} Periodogram of  MORC TTVs alone.  \textit{Bottom Panel:} Periodogram of MORC plus GM15 derived TTVs. We find no evidence for convincing periodic signals in either periodogram. \label{fig:qatar1periodogram}}
\end{center}
\end{figure}

The top panel of Figure \ref{fig:qatar1periodogram} shows the Lomb-Scargle periodogram of our Qatar-1b TTV data. The down-pointing arrows labeled 133 and 267 (epochs) mark the $\sim190$ and $\sim380$~d periods investigated by E13. We find no evidence for those or any other convincing periodic signals in the MORC TTV data.

\begin{figure}
\begin{center}
\resizebox{\columnwidth}{!}{
\includegraphics*{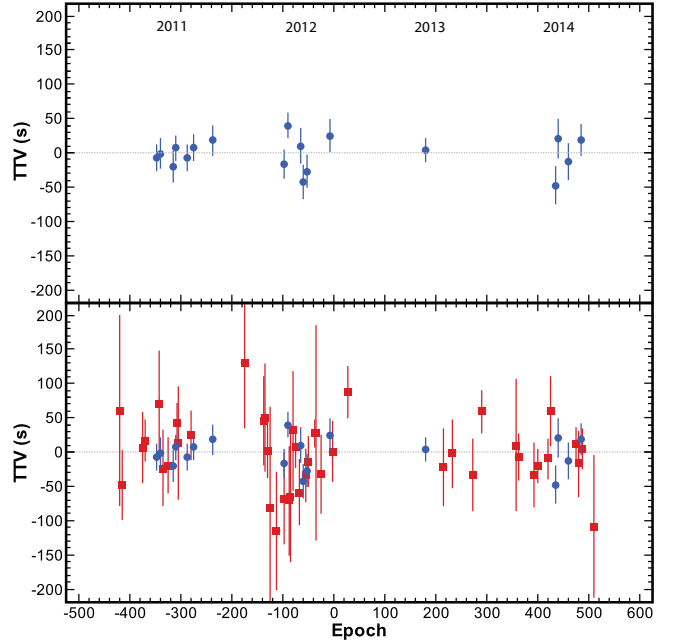} }
\caption[Qatar-1b TTVs]{Qatar-1b TTVs vs. transit epoch. The epoch is calculated from the global fit ephemeris in Table \ref{tab:wasp12fitparmvalues}. TTV is defined as the observed $T_{C}$ minus $T_{C}$ calculated from the linear ephemeris (i.e. $O-C$).  MORC data are shown as circular blue symbols. GM15 derived data are shown as square red symbols. The numbers across the top of the top panel indicate the approximate calender year. \textit{Top Panel:} MORC TTVs alone. \textit{Bottom Panel:} MORC plus  derived TTVs. \label{fig:qatar1ttvslinear}}
\end{center}
\end{figure}

GM15 reanalyzed light curves from E13 and C13 and combined the results with 18 new Qatar-1b transits. We re-phase the GM15 transit center times based on the refined ephemeris from this work (which is nearly identical to the GM15 ephemeris; see Table \ref{tab:qatar1comparison}) and plot the resulting TTVs (square red symbols), along with the TTVs from this work (circular blue symbols), in the bottom panel of Figure \ref{fig:qatar1ttvslinear}. With respect to the linear ephemeris model, $\chisqred=0.82$ for the combined data. It appears that the TTV uncertainties from GM15 are slightly overestimated, and that the linear model is a good representation of the combined data set, unless the uncertainties from this work and GM15 are significantly overestimated. The bottom panel of Figure \ref{fig:qatar1periodogram} shows the Qatar-1b combined data Lomb-Scargle periodogram. The significance of the strongest peak is even less than in the periodogram of our data alone. We find no support for periodic TTVs with semi-amplitude greater than $\sim25$~sec in the Qatar-1b system. 

Even for systems without astrophysical TTVs, if correlated noise systematics are dominant in the data, the TTV scatter from a linear ephemeris should be significantly higher than what would be expected based on the TTV uncertainties extracted by \multifast (see \S \ref{sec:globalfit}). Tables  \ref{tab:wasp12ttvs} and \ref{tab:qatar1ttvs} show that the standard deviations of our TTVs for WASP-12b and Qatar-1b, respectively, are only 6\% larger than the mean of the $1\sigma$ TTV uncertainties. Apparently, correlated noise is minimally impacting our extracted TTV uncertainties, so we expect the same, or even less, impact on the other extracted system parameter uncertainties.  

\section{Summary and Conclusions}\label{sec:wasp12conclusions}

We present the results of self-consistent global fits to 23 WASP-12b light curves and 18 Qatar-1b light curves (that have all been homogeneously reduced starting from the individual images), RV and spectroscopic data from the literature, the \citet{yy04} stellar models, and the \citet{cb11} stellar limb darkening models. We reach space-like photometric precision of 183~ppm and 255~ppm for WASP-12b and Qatar-1b, respectively, per five minute bin in the combined light curve model residuals. 

The WASP-12b derived system parameters in this work are consistent with values from the literature at a level of $1\sigma$, except for the period of the linear ephemeris compared to M13. $M_*$, $R_*$, $M_P$, $R_P$, and $T_{EQ}$ from this work are at the upper end of the $1\sigma$ range of the other studies, while $\loggp$ and $\rho_P$ are at the lower end of the $1\sigma$ range.

The Qatar-1b system parameters derived in this work are consistent with values from the literature at a level of $\sim1\sigma$, except for the initial planetary mass underestimate by A11. The values fall near the mean of the literature values, except $T_{eq}$ from this work is at the upper end of $1\sigma$ relative to the literature values.

The WASP-12b linear ephemeris from this work differs from the M13 ephemeris by $3\sigma$, so a longer baseline of observations may be required to determine the precise ephemeris. Or, if the two earliest TTV measurements shown in Figure \ref{fig:wasp12literaturettvslinear} are accurate, a long term TTV signal (of about $\sim20$~years) may be causing the apparent changes in the shifting linear ephemeris. If the two measurements are found to be inaccurate and are removed from the analysis, the linear ephemeris from this work is a good fit to the combined high-precision timing data from M13 and this work. The $\sim10\%$ uncertainty in the stellar models is not accounted for by multi-EXOFAST, so no attempt has been made to correct for the $\sim1\%$ underestimate of $R_P/R_*$ in the global model. Because of this, our global model parameter values are directly comparable to those in the literature. We do not expect the unaccounted for stellar model uncertainties, nor the blended binary, to affect our TTV measurements. Nevertheless, we aim to improve our global modeling capabilities to account for stellar model uncertainties in future works.

Unless \multifast is overestimating the parameter errors (which is unlikely given the error scaling values listed in Table \ref{tab:wasp12alltransits}), the WASP-12b transit center times from this work are well modeled by a linear ephemeris with $T_{\rm 0}=2456176.668258~\pm$ $7.7650773\times10^{-5}$ and $P=1.0914203~\pm $ $1.4432653\times10^{-7}$, which has $\chisqred=1.09$. A sinusoidal fit to the data using the period corresponding to the highest power peak in the periodogram (3.996 days), yields an improved fit with $\chisqred=0.66$. However, that peak has an analytical false alarm probability of more than 100\%. Our interpretation is that the TTV signal is the result of a night-to-night systematic or a chance fit to noise. Based on the reduced chi-squared value for the linear ephemeris model and the lack of signals in the periodogram, we find no convincing evidence for sinusoidal TTVs with a semi-amplitude of more than $\sim35$~s in the WASP-12b data from this work alone or in combination with data from the literature. 

The Qatar-1b transit center times are also well modeled by a linear ephemeris with $T_{\rm 0}=2456234.103218$ $\pm$ $6.0708415\times10^{-5}$ and $P=1.4200242 $ $\pm$ $2.1728848\times10^{-7}$, which has $\chisqred=1.08$. A Lomb-Scargle periodogram shows no periodic signals in the TTV data that have an analytical false alarm probability less than 100\%. Based on the reduced chi-squared value for the linear ephemeris model and the lack of significant signals in the periodogram, We find no convincing evidence for periodic TTVs with a semi-amplitude of more than $\sim25$~s in the Qatar-1b data. This interpretation is consistent with the conclusion by MG15.

The lack of significant TTVs in these systems is consistent with the \citet{steffen2012} study that found no evidence of TTVs in the orbits of Kepler hot Jupiter planets with $1\le P\le 5$ days. On the other hand, the data are sparsely sampled and it may be possible that short period, low level, or non-sinusoidal TTV signals are lurking in the data. In particular, the WASP-12b timing data between epochs 0 and 200 (the 2012-2013 observing season) seem to show a correlated downward trend. However, the data are too sparse to consider fitting non-sinusoidal TTV signals.

Finally, the results of our extensive new transit observations and systematic re-fitting of the global system parameters has resulted in updated parameters and uncertainties for the WASP-12b and Qatar-1b systems. The WASP-12b system parameter values from this work are consistent with values from previous studies, but typically have $\sim40-50$\% smaller uncertainties than those reported by S12 and C12 as shown in Table \ref{tab:wasp12comparison}. Most of the Qatar-1b system parameter values and uncertainties from this work are consistent with values recently reported by DM15 and GM15 as shown in Table \ref{tab:qatar1comparison}.

\acknowledgments
K.A.C. acknowledges support from NASA Kentucky Space Grant Consortium Graduate Fellowships. 
K.A.C and K.G.S. acknowledge support from NSF PAARE grant AST-1358862 and the Vanderbilt Initiative in Data-intensive Astrophysics.  
We thank the anonymous referee for a thoughtful reading of the manuscript and for useful suggestions. 
This work has made use of NASA's Astrophysics Data System, the Exoplanet Orbit Database at exoplanets.org, the Extrasolar Planet Encyclopedia at exoplanet.eu \citep{epe11}, and the SIMBAD database operated at CDS, Strasbourg, France.
The authors would like to thank Antonio Claret for computing the quadratic limb darkening coefficients for the CBB and Open filter bands which were critical to properly fitting many of the light curves presented herein. K.A.C would like to thank Jason Eastman for insightful TTV and correlated noise discussions and for the development and support of multi-EXOFAST.

\end{document}